\documentclass[aps,pra,reprint,superscriptaddress,longbibliography,nofootinbib,floatfix]{revtex4-2}
\usepackage{amsmath,amssymb,amsthm,mathtools}
\usepackage{graphicx}
\usepackage{bm}
\usepackage{xcolor}
\usepackage[colorlinks=true,linkcolor=blue!55!black,citecolor=blue!55!black,urlcolor=blue!55!black]{hyperref}

\newtheorem{theorem}{Theorem}
\newtheorem{lemma}{Lemma}
\newtheorem{proposition}{Proposition}
\newtheorem{corollary}{Corollary}

\newcommand{\ket}[1]{\lvert #1\rangle}
\newcommand{\bra}[1]{\langle #1\rvert}
\newcommand{\braket}[2]{\langle #1\vert #2\rangle}
\newcommand{\ketbra}[2]{\lvert #1\rangle\langle #2\rvert}
\newcommand{\cT}{\mathcal{T}}
\newcommand{\cN}{\mathcal{N}}
\newcommand{\cX}{\mathcal{X}}
\newcommand{\vphi}{{\bm{\varphi}}}
\newcommand{\vth}{{\bm{\vartheta}}}
\newcommand{\Pst}{P^{\star}}
\newcommand{\lmin}{\lambda_{\min}}
\newcommand{\lmax}{\lambda_{\max}}
\newcommand{\tr}{\operatorname{Tr}}
\newcommand{\diag}{\operatorname{diag}}
\newcommand{\one}{\openone}
\newcommand{\ones}{\bm{1}}
\newcommand{\Sym}{\mathfrak{S}}

\begin{document}

\title{Probabilistic distillation of superposition}

\author{G\"{o}khan Torun}
\email{gtorung@gmail.com}
\affiliation{Department of Physics, Faculty of Arts and Sciences, \.{I}zmir University of Economics, 35330 \.{I}zmir, T\"{u}rkiye}

\date{\today}

\begin{abstract}
How well a superposition of nonorthogonal states, such as an optical cat state, serves as a resource depends on how its components interfere. Here we determine the optimal probability of converting one such superposition into another by superposition-free operations. These operations cannot make the components of a superposition interfere more destructively with certainty while keeping all of them, and this principle fixes the optimal probability for every source that is symmetric under permutations of its components. When the reference states are coherent states, as in a cat state, every operation that maps each of them to a coherent state, such as photon loss or photon subtraction, is superposition-free. For two reference states we solve the problem completely. Superpositions with coefficients of equal modulus, the counterparts of the maximally coherent states of coherence theory, are therefore inequivalent resources: an odd cat state becomes even with certainty by a normalized photon subtraction, whereas an even one becomes odd only with a probability that vanishes for small amplitudes, and the even and the Yurke--Stoler cat states cannot be converted into each other with certainty in either direction. Collective operations on several copies can do better than converting them one at a time, except in the distillation of coherence, which gains nothing from them. For three or more reference states with equal positive overlaps, no state reaches every other state with certainty; the equal-modulus superpositions that interfere most destructively still yield a maximally coherent state with certainty and reach every superposition of all reference states at least as well as such a state does, but in the worst case no better.
\end{abstract}

\maketitle

%=====================================================================
\section{Introduction}\label{sec:intro}
%=====================================================================

Superposition is a defining nonclassical feature of quantum mechanics and a resource for quantum technologies~\cite{Chitambar2019}. For orthonormal reference states it is quantified by the resource theory of coherence~\cite{Aberg2006,Baumgratz2014,Winter2016,Streltsov2017}, in which a single family of states plays all distinguished roles: the maximally coherent states, the superpositions of all basis states with coefficients of equal modulus, have equal populations, are related to one another by free unitaries~\cite{Peng2016}, reach every state with certainty~\cite{Baumgratz2014}, and are the targets of distillation~\cite{Winter2016,Regula2018,TorunLami2019,LiuZhou2019}. When certainty is abandoned, the figure of merit is the optimal success probability of a single-shot conversion. It is known in closed form for pure entangled states~\cite{Vidal1999} and for pure states in coherence theory~\cite{Fang2018,LiuSun2021}, and general limits on probabilistic transformations hold in every resource theory~\cite{FangLiu2020,Regula2022}.

Yet many physically relevant reference states are not orthogonal. Coherent states \(\ket\alpha\)~\cite{Glauber1963} are the most classical pure states of light and of other bosonic modes, and their superpositions, the cat states, are resources for continuous-variable quantum information and bosonic error correction~\cite{Dodonov1974,Ourjoumtsev2006,Vlastakis2013,Mirrahimi2014,Grimsmo2020} that can be converted into entanglement~\cite{Killoran2016}; overlapping atomic orbitals~\cite{Yesiller2024} and the signal states of unambiguous discrimination~\cite{Chefles1998,CheflesBarnett1998} are further examples. The resource theory of superposition (RTS)~\cite{Theurer2017} captures these settings; related frameworks build on positive-operator-valued measures or biorthogonal bases~\cite{Bischof2019,Srivastava2020,Pusuluk2022}. Its free states are mixtures of linearly independent reference states, and the Kraus operators of its free operations, the superposition-free operations (FO), map each reference state to a multiple of a reference state. For coherent states as reference states, every operation that maps each of them to a coherent state, such as a phase-space rotation, a displacement, photon loss, or photon subtraction, is superposition-free, with the images as the new reference states (Sec.~\ref{sec:FO}). Optimal probabilities under superposition-free operations therefore limit all such operations, and RTS with finitely many coherent states as reference states is a finite-dimensional counterpart of the resource theories of nonclassicality, whose free states are the mixtures of coherent states~\cite{Tan2017,Yadin2018}.

Nonorthogonality breaks the unity of the maximally coherent states. The even and odd cat states, the superpositions of the coherent states \(\ket\alpha\) and \(\ket{-\alpha}\) with equal and with opposite signs (Fig.~\ref{fig:qubit_geometry}), have coefficients of equal modulus, exactly like the states \(\ket\pm\) of coherence theory, but they are not interchangeable. That the subtraction or loss of a photon turns each of them into the other is well known: it is used to generate optical cat states~\cite{Dakna1997,Ourjoumtsev2006} and is the dominant error of cat codes~\cite{Mirrahimi2014,Grimsmo2020}, and small cat states can be enlarged probabilistically~\cite{Lund2004,Sychev2017}. The optimal probabilities of such conversions, however, have to our knowledge not been determined. We show that they are fixed by interference: superposition-free operations cannot make the components of a superposition interfere more destructively with certainty while keeping all of them (Theorem~\ref{thm:inphase}). The components of the odd cat state interfere destructively and those of the even cat state constructively, so the odd cat state becomes even with certainty, by a normalized photon subtraction, whereas the even cat state becomes odd at best with probability \(\tanh^2|\alpha|^2\), the ratio of their mean photon numbers [Fig.~\ref{fig:graphs}(a)].

Within RTS, deterministic conversions obey overlap-dependent majorization-type conditions~\cite{Torun2021}, which Ref.~\cite{Torun2026b} cast as Gram majorization, a majorization relation between the Gram matrices of the components of two states. Golden states, which reach every state with certainty under FO, exist only for special overlaps~\cite{Senyasa2022,Torun2026b}. For probabilistic conversions, Theurer \emph{et al.} showed that the success probability is nonzero iff the number of nonzero coefficients does not increase, and that for a target with as many nonzero coefficients as the source the optimal probability is the value of a semidefinite program~\cite{Theurer2017}. Closed forms for this probability, the optimal protocols, and an answer to the question of which superpositions with coefficients of equal modulus are the better resources have not been available; probabilistic distillation was raised as an open direction in Refs.~\cite{Senyasa2022,Torun2023}.

Our organizing tool is L\"owdin symmetric orthogonalization~\cite{Lowdin1950,Carlson1957,Aiken1980,Mayer2002}, which Ref.~\cite{Torun2023} introduced into the resource theories of coherence and superposition. It produces the orthonormal basis closest to the reference states, the L\"owdin frame. Its inverse maps the maximally coherent states of this frame onto the states whose superposition coefficients have equal moduli~\cite{Torun2023}, which we call \emph{balanced superposition states}, and it turns the Kraus operators of incoherent operations into superposition-free Kraus operators. The question is how much of coherence theory survives this crossing.

We obtain two sets of results. The first concerns arbitrary reference states. The optimal probability is bounded from above by the ratio of two numbers that measure how destructively the components of the source and of the target interfere, and from below by the probability of a route through a maximally coherent state. The upper bound is attained whenever the Gram matrix of the components of the source is invariant under permutations, and the optimum is a single generalized eigenvalue whenever that of the target is (Theorem~\ref{thm:inphase}). These cases come with explicit optimal Kraus operators and contain the even and odd cat states, the maximally coherent states, and the in-phase state, the superposition of equal-overlap reference states with equal coefficients. For several copies, collective operations turn even cat states into odd ones with a higher probability than converting the copies one at a time (Proposition~\ref{prop:copies}), whereas the distillation of coherence from superposition gains nothing from them (Corollary~\ref{cor:copies}). For two reference states we solve the problem for every source and target (Proposition~\ref{prop:qubit}). This gives all conversions among cat states with an arbitrary relative phase, such as the Yurke--Stoler cat states generated by a Kerr nonlinearity~\cite{Yurke1986}, and shows that the even and the Yurke--Stoler cat states are incomparable (Corollary~\ref{cor:phase}).

The second set concerns the balanced states, which we characterize for equal overlaps \(s\) in every dimension \(d\) (Theorem~\ref{thm:images}). Unlike the maximally coherent states, they are inequivalent resources (Fig.~\ref{fig:graphs}), and for \(d\ge3\) the sign of \(s\) decides which roles of the maximally coherent states survive (Table~\ref{tab:roles}). For \(s<0\) a single state, the in-phase state, reaches every state with certainty~\cite{Senyasa2022}. For \(s>0\) the interference of three or more components is frustrated~\cite{Torun2026b}, and no state does: even the balanced states whose components interfere most destructively fail for balanced targets, and for three reference states we give their optimal probability for every target with three nonzero coefficients (Theorem~\ref{thm:nogo}). They nevertheless yield a maximally coherent state with certainty, through an unambiguous discrimination of the reference states that never fails on them (Theorem~\ref{thm:bridge}), and they reach every state with all coefficients nonzero at least as well as a maximally coherent state does; this guarantee, the price of frustration, is exact in the worst case (Theorem~\ref{thm:price}).

The remainder of the paper is organized as follows. Section~\ref{sec:setting} introduces the setting, Sec.~\ref{sec:images} the balanced states and their minimal weight, Sec.~\ref{sec:sdp} the optimal probability, with the results for qubits and cat states, and Sec.~\ref{sec:nogo} the inequivalence of the balanced states; Sec.~\ref{sec:conclusion} summarizes the results and lists open questions, and the appendixes contain the proofs and the numerical checks.

%=====================================================================
\section{Setting}\label{sec:setting}
%=====================================================================

\subsection{Reference states, dual basis, and L\"owdin frame}

Let the reference states \(\{\ket{c_j}\}_{j=0}^{d-1}\) be normalized and linearly independent, forming a basis of \(\mathcal H\cong\mathbb C^d\), with Gram matrix \(S\) of overlaps \(S_{jk}=\braket{c_j}{c_k}\), \(S=S^\dagger>0\). Every pure state is a superposition of the reference states,
\begin{equation}
\ket\psi=\sum_jx_j\ket{c_j},\quad x^\dagger Sx=1,
\label{eq:expansion}
\end{equation}
with superposition coefficients \(x_j=\braket{c_j^\perp}{\psi}\)~\cite{Torun2021,Senyasa2022}, which form the coefficient vector \(x=(x_0,\dots,x_{d-1})^T\). A state, or its coefficient vector, has \emph{full support} if all its coefficients are nonzero. Here \(\{\ket{c_j^\perp}\}\) is the dual basis, \(\ket{c_j^\perp}=\sum_k(S^{-1})_{kj}\ket{c_k}\), which satisfies \(\braket{c_j^\perp}{c_k}=\delta_{jk}\)~\cite{Theurer2017,Senyasa2022}. We call \(|x_j|^2\) the \emph{weights} of \(\ket\psi\); in coherence theory they are the populations, but in general they do not sum to one. Unless stated otherwise, \(S\) is arbitrary; equal overlaps are assumed only in Sec.~\ref{sec:equal} and where indicated.

L\"owdin symmetric orthogonalization constructs from the nonorthogonal reference states the \emph{L\"owdin frame} \(\{\ket{l_j}\}\)~\cite{Lowdin1950}, the orthonormal basis closest to them: it minimizes \(\sum_j\|\ket{c_j}-\ket{l_j}\|^2\) over all orthonormal bases~\cite{Carlson1957,Aiken1980,Mayer2002}. The orthonormal L\"owdin vectors and the nonorthogonal reference states are related by
\begin{equation}
\ket{l_j}=\sum_k(S^{-1/2})_{kj}\ket{c_k},\quad\ket{c_j}=\sum_k(S^{1/2})_{kj}\ket{l_k}.
\label{eq:lowdin}
\end{equation}
The \emph{inverse L\"owdin map}
\begin{equation}
\cT=\sum_j\ketbra{c_j}{l_j},\quad[\cT]_{\rm L}=S^{1/2},
\label{eq:Tdef}
\end{equation}
with \([\,\cdot\,]_{\rm L}\) the matrix in the L\"owdin frame, sends \(\ket{l_j}\mapsto\ket{c_j}\): it turns the L\"owdin coefficients of a state into the superposition coefficients of its image. Its inverse, the \emph{forward L\"owdin map}, sends \(\ket{c_j}\mapsto\ket{l_j}\) and \(\ket{l_j}\mapsto\ket{c_j^\perp}\). Unless \(S=\one\), neither map is unitary: \(\|\cT\|^2=\lmax(S)>1\) and \(\|\cT^{-1}\|^2=1/\lmin(S)>1\), so as Kraus operators they must be rescaled (Theorem~\ref{thm:bridge}).

Two objects attached to \(\ket\psi\) organize the analysis: the \emph{coefficient operator}, which maps each \(\ket{c_j}\) to \(x_j\ket{c_j}\), and the \emph{component Gram matrix},
\begin{equation}
\cX_\psi=\sum_jx_j\ketbra{c_j}{c_j^\perp},\quad G_\psi=X^\dagger SX,
\label{eq:Gpsi}
\end{equation}
with \(X=\diag(x)\), so that \(G_\psi\) is the Gram matrix of the components \(x_j\ket{c_j}\) of \(\ket\psi\) and carries the weights \(|x_j|^2\) on its diagonal; in coherence theory (\(S=\one\)) it is the diagonal matrix of the populations. For \(\ket u=\sum_jy_j\ket{c_j}\),
\begin{equation}
\|\cX_\psi\ket u\|^2=y^\dagger G_\psi\,y,
\label{eq:Gform}
\end{equation}
because \(\cX_\psi\ket u=\sum_jx_jy_j\ket{c_j}\) has the coefficient vector \(Xy\). The component Gram matrix was introduced in Ref.~\cite{Torun2026b}, where it decides deterministic conversions. Its trace is the squared Euclidean norm \(\|x\|^2=\sum_j|x_j|^2\), and the sum of its entries is \(\ones^TG_\psi\ones=\braket{\psi}{\psi}=1\), with \(\ones=(1,\dots,1)^T\). We use the normalized squared norm
\begin{equation}
\tau(\psi)=\frac{d\,\|x\|^2-1}{d-1},
\label{eq:tau}
\end{equation}
which is positive because \(\|x\|^2\ge1/\lmax(S)>1/d\)~\cite{Senyasa2022}. Since \(x^\dagger Sx=1\), the difference \(\|x\|^2-1=-2\,{\rm Re}\sum_{j<k}\bar x_jS_{jk}x_k\) is minus the interference term of the components: \(\tau(\psi)>1\) when they interfere destructively, \(\tau(\psi)<1\) when they interfere constructively, and \(\tau=1\) in coherence theory, where components do not interfere.

\subsection{Superposition-free operations}\label{sec:FO}

A Kraus operator is superposition-free iff it maps each reference state to a multiple of a reference state~\cite{Theurer2017}, that is, iff
\begin{equation}
K=\sum_k\xi_k\ketbra{c_{h(k)}}{c_k^\perp}
\label{eq:FOkraus}
\end{equation}
for coefficients \(\xi_k\) and an index map \(h\). Superposition-free operations (FO) are the channels whose Kraus operators can all be chosen of this form~\cite{Theurer2017,Torun2026b}; they form a proper subset of the maximal superposition-free operations, which only have to map free states to free states~\cite{Theurer2017}. In the L\"owdin frame, the Kraus operator~\eqref{eq:FOkraus} reads
\begin{equation}
[K]_{\rm L}=S^{1/2}\,\Pi_hD_\xi\,S^{-1/2},
\label{eq:lowdinFO}
\end{equation}
with \((\Pi_h)_{jk}=\delta_{j,h(k)}\) and \(D_\xi=\diag(\xi)\), and the completeness relation becomes
\begin{equation}
\textstyle\sum_nK_n^\dagger K_n=\one\;\Longleftrightarrow\;\sum_nD_{\xi_n}^\dagger\Pi_{h_n}^TS\,\Pi_{h_n}D_{\xi_n}=S.
\label{eq:lowdinFOcomplete}
\end{equation}
The Kraus operators of FO are therefore incoherent Kraus operators~\cite{Baumgratz2014} of the L\"owdin frame conjugated by the inverse L\"owdin map, with completeness measured in the metric \(S\). A stochastic FO is a set \(\{K_n\}\) of such operators with \(\sum_nK_n^\dagger K_n\le\one\). The optimal probability of converting \(\ket\psi\) (coefficients \(x\)) into \(\ket\phi\) (coefficients \(t\)) is
\begin{equation}
\Pst(\psi\to\phi)=\sup\textstyle\sum_n\|K_n\ket\psi\|^2,
\label{eq:Pdef}
\end{equation}
the supremum being over stochastic FO \(\{K_n\}\) with \(K_n\ket\psi\propto\ket\phi\) for all \(n\). We call \(\ket\psi\) the source and \(\ket\phi\) the target. \emph{Superposition distillation} is the case of a balanced target (Sec.~\ref{sec:balanced}). We take \(t\) with full support; then \(\Pst=0\) unless \(x\) also has full support~\cite{Theurer2017}, so we take \(x\) with full support as well. For a permutation \(\pi\), the operator \(U_\pi=\sum_k\ketbra{c_{\pi(k)}}{c_k^\perp}\) permutes the reference states, and \((\Pi_\pi)_{jk}=\delta_{j,\pi(k)}\).

\begin{lemma}[Branch structure]
\label{lem:branches}
(a) If \(K\ket\psi=z\ket\phi\neq0\) for a superposition-free Kraus operator \(K\), then the index map \(h\) of \(K\) in Eq.~\eqref{eq:FOkraus} is a permutation \(\pi\) and \(K=z\,\cX_\phi U_\pi\cX_\psi^{-1}\).
(b) Superposition-free operators with \(\sum_nK_n^\dagger K_n\le\one\) are completed to a trace-preserving FO by the operators \(L_\kappa=\ketbra{c_0}{\kappa}\), where the kets \(\ket\kappa\) are those of any decomposition \(\one-\sum_nK_n^\dagger K_n=\sum_\kappa\ketbra{\kappa}{\kappa}\).
(c) For equal real overlaps \(S_{jk}=s\neq0\) (\(j\neq k\)) the unitary FO, or free unitaries, are the operators \(U_\pi\), up to a global phase.
\end{lemma}

Every successful branch therefore \emph{unwinds} the source, \(\cX_\psi^{-1}\ket\psi=\sum_j\ket{c_j}\), \emph{permutes} the reference states, and \emph{imprints} the target coefficients, while the failure branches can be taken to output the free state \(\ket{c_0}\). By (c), for equal real overlaps \(s\neq0\) two states whose coefficients have equal moduli are related by a free unitary only if their phase patterns differ by a permutation and a global phase~\cite{Theurer2017,Senyasa2022}, in contrast to coherence theory.

The same definition applies to Kraus operators \(K:\mathcal H\to\mathcal H'\) between spaces with reference states \(\{\ket{c_j}\}\) and \(\{\ket{c'_k}\}\) (Gram matrices \(S\) and \(S'\)): \(K\) is superposition-free if \(K\ket{c_j}\propto\ket{c'_{h(j)}}\) for all \(j\). We call operations with such Kraus operators \emph{superposition-free maps} from the first reference set to the second. They map free states to free states and are closed under composition, so that no sequence of them creates superposition. For identical reference sets they are the FO, and for orthonormal reference states on both sides they are the incoherent operations of coherence theory~\cite{Baumgratz2014}; with nonorthogonal reference states on one side and an orthonormal basis on the other they connect the two theories. The L\"owdin maps are the coefficient-preserving members of this class: \(\cT\) is superposition-free from the L\"owdin frame to \(\{\ket{c_j}\}\), and \(\cT^{-1}\) in the opposite direction. When the source and the target refer to different reference sets, \(\Pst\) in Eq.~\eqref{eq:Pdef} is the supremum over the superposition-free maps between them. Lemma~\ref{lem:branches}(a),(b), Proposition~\ref{prop:mu}(a),(b), and Theorems~\ref{thm:sdp} and~\ref{thm:inphase} below then hold unchanged for reference sets of equal size, with \(G_\phi\) evaluated with \(S'\) and with \(\ket{c_0}\) in Lemma~\ref{lem:branches}(b) replaced by \(\ket{c'_0}\).

Let the reference states be coherent states \(\ket{c_j}=\ket{\alpha_j}\) of a bosonic mode, for instance a mode of light or of a microwave cavity. Then phase-space rotations, displacements, photon loss, and (suitably normalized) photon subtraction, restricted to the span of the reference states, are superposition-free maps, with the images of the \(\ket{\alpha_j}\) as the reference states of the output. Indeed, with the annihilation operator \(\hat a\) and the displacement operator \(\hat D(\beta)\), \(e^{-i\Theta\hat a^\dagger\hat a}\ket\alpha=\ket{\alpha e^{-i\Theta}}\), \(\hat D(\beta)\ket\alpha=e^{i\,{\rm Im}(\beta\bar\alpha)}\ket{\alpha+\beta}\), and \(\hat a\ket\alpha=\alpha\ket\alpha\); and a beam splitter of transmissivity \(\eta>0\) maps \(\ket\alpha\ket0\) to a product \(\ket{\sqrt\eta\,\alpha}\ket{\alpha'}\) of coherent states, with \(|\alpha'|^2=(1-\eta)|\alpha|^2\), so that every Kraus operator of the resulting loss channel maps \(\ket\alpha\) to a multiple of \(\ket{\sqrt\eta\,\alpha}\). More generally, a channel that maps each reference state to a coherent state is a superposition-free map, because each of its Kraus operators must then map that reference state to a multiple of the same coherent state.

\subsection{Balanced superposition states}\label{sec:balanced}

Let \(\ket{\Phi_\vphi}=d^{-1/2}\sum_je^{i\varphi_j}\ket{l_j}\), \(\varphi_0=0\), be the maximally coherent states of the L\"owdin frame. Their normalized inverse L\"owdin images are
\begin{equation}
\ket{\Psi_\vphi}=\frac{\cT\ket{\Phi_\vphi}}{\|\cT\ket{\Phi_\vphi}\|}=\frac{1}{\sqrt{d\,\cN(\vphi)}}\sum_je^{i\varphi_j}\ket{c_j},
\label{eq:BSS}
\end{equation}
where \(\cN(\vphi)=\tfrac1de_\vphi^\dagger Se_\vphi\) and \((e_\vphi)_j=e^{i\varphi_j}\); we write \(\cN\) for \(\cN(\vphi)\) where the pattern is evident. Reference~\cite{Torun2023} identified these images, under the name maximal superposition states; the \emph{in-phase state} \(\ket{\Psi_0}\) (\(\vphi=0\)) is the state \(\ket{\Psi_d^+}\) of Refs.~\cite{Senyasa2022,Torun2023}. The states~\eqref{eq:BSS} are, up to a global phase, exactly the states whose superposition coefficients have equal moduli: \(\cT\) carries the L\"owdin coefficients \(e^{i\varphi_j}/\sqrt d\) unchanged into superposition coefficients, up to the factor \(\cN(\vphi)^{-1/2}\), whereas maximal coherence in the L\"owdin frame survives only for the phase patterns \(\vphi\) of Theorem~\ref{thm:images} below. We call the states~\eqref{eq:BSS} \emph{balanced superposition states}, or balanced states for short; their weights \(|x_j|^2=1/[d\,\cN(\vphi)]\) are equal.

Three properties that coincide in coherence theory, namely equal weights, the largest minimal weight, and the ability to reach every state with certainty, no longer coincide for nonorthogonal reference states. The minimal weight of Sec.~\ref{sec:images} varies across the balanced family, and balanced and golden states are different notions: the first is defined by the coefficients, the second by an operational property, the ability to reach every state with certainty under FO~\cite{Senyasa2022,Torun2026b}. For equal overlaps the golden states are balanced: a golden state reaches in particular the states of largest minimal weight, so it has the largest minimal weight itself by Proposition~\ref{prop:mu}(b), and such states are balanced by Proposition~\ref{prop:mu}(c). The converse fails: for \(s<0\) the in-phase state is the only golden state, and for \(0<s<1\) and \(d\ge3\) no state is golden~\cite{Torun2026b} (Table~\ref{tab:roles}).

\subsection{Equal overlaps}\label{sec:equal}

Our explicit results concern equal real overlaps,
\begin{equation}
S=(1-s)\one+sJ,\quad-\tfrac{1}{d-1}<s<1,
\label{eq:Sequal}
\end{equation}
with \(J=\ones\ones^T\)~\cite{Torun2021,Senyasa2022,Torun2023}. The spectrum of \(S\) is
\begin{equation}
\lambda_0=1+(d-1)s\ \text{on}\ \ones,\quad\lambda_1=1-s\ \text{on}\ \ones^\perp.
\label{eq:spectrum}
\end{equation}
In the L\"owdin frame these eigenspaces are spanned by the Fourier states \(\ket{f_k}=d^{-1/2}\sum_j\omega^{jk}\ket{l_j}\), \(\omega=e^{2\pi i/d}\), the discrete Fourier basis of the L\"owdin frame, with \(k=0\) and \(k\neq0\), respectively, and, with \(p=\sqrt{\lambda_0}\) and \(q=\sqrt{\lambda_1}\),
\begin{equation}
\cT=q\,\one+(p-q)\ketbra{f_0}{f_0},\quad\ket{c_j}=q\ket{l_j}+\frac{p-q}{\sqrt d}\ket{f_0}.
\label{eq:Tequal}
\end{equation}
Each \(\ket{c_j}\) is thus the L\"owdin vector rescaled by \(q\) (shrunk for \(s>0\)) plus a common component along \(\ket{f_0}\). Since \(\cT\) is diagonal in the Fourier basis, \(\ket{f_0}=\ket{\Psi_0}\), and each \(\ket{f_k}\) is the balanced state with phases \(\varphi_j=2\pi jk/d\). For \(d=3\), the states \(\ket{f_1}\) and \(\ket{f_2}\) are the candidates for golden states that were examined, and excluded for \(s=\tfrac12\), in Ref.~\cite{Theurer2017}.

Because \(1/\lmax(S)\le\|x\|^2\le1/\lmin(S)\), the normalized squared norm~\eqref{eq:tau} lies between the values
\begin{equation}
\tau_0=\frac{\lambda_1}{\lambda_0},\quad\tau_A=\frac{d-1+s}{(d-1)\lambda_1},
\label{eq:range}
\end{equation}
which it takes for \(x\propto\ones\), the in-phase state, and for every \(x\perp\ones\), respectively: \(\tau_0\le\tau(\psi)\le\tau_A\) for \(s>0\), and \(\tau_A\le\tau(\psi)\le\tau_0\) for \(s<0\). Where the overlap varies we write \(\tau_0(s)\) and \(\tau_A(s)\); \(\tau_0\) decreases and \(\tau_A\) increases with \(s\), and both equal one at \(s=0\).

\begin{figure*}[t]
\centering
\includegraphics[width=\textwidth]{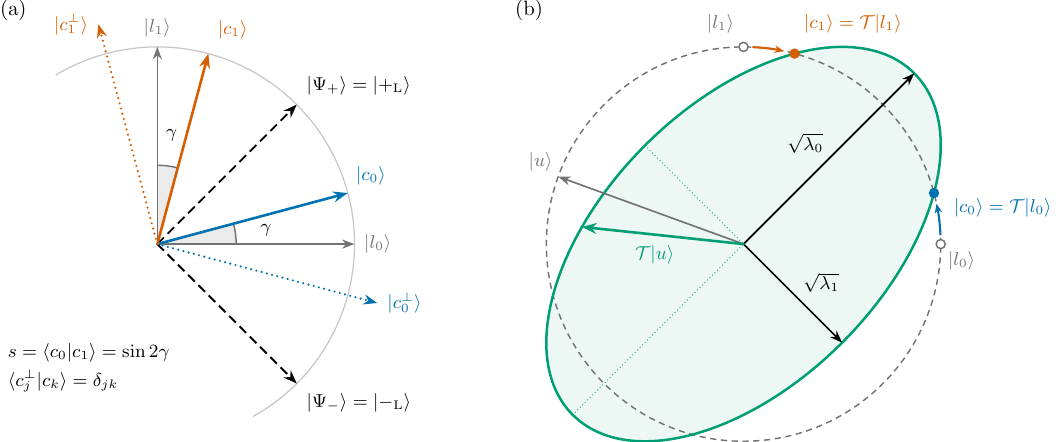}
\caption{Qubit geometry of the inverse L\"owdin map for \(s=1/2\) (\(\gamma=15^\circ\)), drawn from \(S^{1/2}\) in the L\"owdin frame. (a) Reference states \(\ket{c_{0,1}}\), dual vectors \(\ket{c_{0,1}^\perp}\) (dotted, of norm \(1/\sqrt{1-s^2}\); \(\ket{c_1^\perp}\perp\ket{c_0}\) and \(\ket{c_0^\perp}\perp\ket{c_1}\)), L\"owdin frame \(\ket{l_{0,1}}\), and the real balanced states \(\ket{\Psi_\pm}\propto\ket{c_0}\pm\ket{c_1}\) (dashed), which coincide with \(\ket{\pm_{\rm L}}\). (b) \(\cT\) maps the unit circle (dashed) onto an ellipse (green) with semi-axes \(\sqrt{\lambda_0}=\sqrt{1+s}\) along \(\ket{+_{\rm L}}\) and \(\sqrt{\lambda_1}=\sqrt{1-s}\) along \(\ket{-_{\rm L}}\) (dotted lines: principal axes). The images \(\ket{c_j}=\cT\ket{l_j}\) lie on both curves; \(\ket{\pm_{\rm L}}\) keep their direction, while a generic state \(\ket u\) is turned toward the \(\ket{+_{\rm L}}\) axis, and its image \(\cT\ket u\) (green) is not normalized, because \(\cT\) is not unitary. For \(s<0\), \(\gamma<0\): the reference states lean toward the \(\ket{-_{\rm L}}\) axis, and the long axis of the ellipse lies along \(\ket{-_{\rm L}}\). For the coherent states \(\ket{c_{0,1}}=\ket{\pm\alpha}\) one has \(s=e^{-2|\alpha|^2}\), and \(\ket{\Psi_\pm}\) are the even and odd cat states.}
\label{fig:qubit_geometry}
\end{figure*}

Figure~\ref{fig:qubit_geometry} shows the qubit, with \(\ket{\pm_{\rm L}}=(\ket{l_0}\pm\ket{l_1})/\sqrt2\): \(\ket{c_0}\) and \(\ket{c_1}\) lean toward \(\ket{+_{\rm L}}\) by the angle \(\gamma=\tfrac12\arcsin s\), and the real balanced states \(\ket{\Psi_\pm}\propto\ket{c_0}\pm\ket{c_1}\) coincide with \(\ket{\pm_{\rm L}}\); thus \(\ket{\Psi_+}=\ket{\Psi_0}=\ket{f_0}\) and \(\ket{\Psi_-}=\ket{f_1}\). For \(d=2\) the sign of \(s\) is a phase convention: the sign change \(\ket{c_1}\to-\ket{c_1}\) leaves the free states and the FO unchanged, maps \(s\to-s\), and exchanges \(\ket{\Psi_+}\) and \(\ket{\Psi_-}\). Every qubit statement made below for \(s>0\) therefore holds for \(s<0\) after the replacements \(s\to|s|\) and \(\ket{\Psi_\pm}\to\ket{\Psi_\mp}\).

For \(d\ge3\) the sign of \(s\) is physical: the triple product \(S_{01}S_{12}S_{20}=s^3\) is a Bargmann invariant~\cite{Bargmann1964}, unchanged by phase changes of the reference states. Positive overlaps frustrate the interference~\cite{Torun2026b}: three phases cannot differ pairwise by \(\pi\), so no phase pattern lets every pair of components interfere maximally destructively. Shor's lifted trine states~\cite{Shor2004}, obtained by lifting the three planar trine states out of their plane, realize both signs: their common overlap grows from \(-\tfrac12\) in the plane through zero, where they are orthonormal, to positive values. Table~\ref{tab:roles} shows how the distinguished states depend on the sign in every dimension.

%=====================================================================
\section{Balanced states and the minimal weight}\label{sec:images}
%=====================================================================

This section determines which balanced states keep the maximal coherence of their L\"owdin preimages and introduces the minimal weight, which bounds every conversion. Let \(\zeta(\vphi)=\braket{f_0}{\Phi_\vphi}=\tfrac1d\sum_je^{i\varphi_j}\) be the mean phasor. We call a state \emph{bi-balanced} if it is balanced in \(\{\ket{c_j}\}\) and maximally coherent in the L\"owdin frame. The following theorem describes what \(\cT\) does to every maximally coherent state of the L\"owdin frame.

\begin{theorem}[Inverse L\"owdin images]
\label{thm:images}
For equal overlaps \(s\neq0\):
(i)~\(\cN(\vphi)=\lambda_1+ds|\zeta(\vphi)|^2=1+\tfrac{2s}{d}\sum_{j<k}\cos(\varphi_j-\varphi_k)\), and
\begin{equation}
\braket{l_j}{\Psi_\vphi}=\frac{q\,e^{i\varphi_j}+(p-q)\,\zeta(\vphi)}{\sqrt{d\,\cN(\vphi)}}.
\label{eq:Lcoeff}
\end{equation}
(ii)~\(\ket{\Psi_\vphi}=\ket{\Phi_\vphi}\) iff (A)~\(\zeta(\vphi)=0\), in which case \(\cT\ket{\Phi_\vphi}=q\ket{\Phi_\vphi}\), or (B)~all phases are equal, in which case \(\cT\ket{\Phi_\vphi}=p\ket{\Phi_\vphi}\).
(iii)~\(\ket{\Psi_\vphi}\) is bi-balanced iff (A), (B), or (C) holds, where (C) means that \(d\) is even and the phases take two values, each on \(d/2\) indices, with relative phase \(\varphi\notin\{0,\pi\}\). In case (C), \(\ket{\Psi_\vphi}\) is, up to a global phase, the maximally coherent state with the same two-block pattern and the relative phase \(\varphi'\) given by the phase-contraction law
\begin{equation}
\tan\frac{\varphi'}{2}=\sqrt{\frac{1-s}{1+(d-1)s}}\,\tan\frac{\varphi}{2}.
\label{eq:contraction}
\end{equation}
\end{theorem}

\begin{table}[b]
\caption{Bi-balanced states for equal overlaps \(s\neq0\) (Theorem~\ref{thm:images}), given by their phase patterns \(e_\vphi\) up to permutations and a global phase; class (B) is \(\ket{\Psi_0}\) in every dimension.}
\label{tab:classes}
\begin{ruledtabular}
\begin{tabular}{c p{3.45cm} p{3.15cm}}
\(d\) & \raggedright class (A), \(\cT\)-invariant & \raggedright class (C), rotated\tabularnewline
\hline
2 & \raggedright \(\ket{\Psi_-}\) & \raggedright all balanced states except \(\ket{\Psi_\pm}\)\tabularnewline
3 & \raggedright \(\ket{f_1},\ket{f_2}\) & \raggedright none (\(d\) odd)\tabularnewline
4 & \raggedright three circles, \((1,e^{i\varphi},-1,-e^{i\varphi})\) & \raggedright three circles, \((1,1,e^{i\varphi},e^{i\varphi})\) with \(\varphi\notin\{0,\pi\}\)\tabularnewline
\(\ge5\) & \raggedright \((d-3)\)-dimensional family & \raggedright present iff \(d\) even\tabularnewline
\end{tabular}
\end{ruledtabular}
\end{table}

We call the bi-balanced states of classes (A) and (B) \emph{\(\cT\)-invariant}, because \(\cT\) maps \(\ket{\Phi_\vphi}\) to a multiple of itself, and those of class (C) \emph{rotated}, because \(\cT\) maps \(\ket{\Phi_\vphi}\) to a multiple of a different maximally coherent state; by Eq.~\eqref{eq:contraction}, the rotation contracts the relative phase toward \(0\) for \(s>0\) and toward \(\pi\) for \(s<0\). The structure depends on \(d\) (Table~\ref{tab:classes}). For \(d=2\) every balanced state is bi-balanced, and only \(\ket{\Psi_\pm}\) are \(\cT\)-invariant. For \(d=3\), \(\zeta=0\) forces the three phasors \(e^{i\varphi_j}\) to form an equilateral triangle, so class (A) consists of the Fourier states \(\ket{f_1}\) and \(\ket{f_2}\); every other balanced state of the qutrit except \(\ket{\Psi_0}\), for instance the real state \(\ket R=\ket{\Psi_{(0,\pi,\pi)}}\), is not maximally coherent in the L\"owdin frame. For \(d\ge4\) class (A) is a continuum, which contains the Fourier states \(\ket{f_k}\), \(k\neq0\), but also non-Fourier states.

The \emph{minimal weight}
\begin{equation}
\mu(\psi)=\min_j|x_j|^2,
\label{eq:mu}
\end{equation}
the analog of the minimal population in coherence theory~\cite{Fang2018}, connects this structure to distillation. For balanced states \(\mu(\Psi_\vphi)=1/[d\,\cN(\vphi)]\), so by Theorem~\ref{thm:images}(i) it is largest on class (A) for \(s>0\) and on \(\ket{\Psi_0}\) for \(s<0\).

\begin{proposition}[Minimal weight]
\label{prop:mu}
For any Gram matrix \(S\):
(a)~\(\mu\) is a strong monotone: \(\sum_nP_n\mu(\psi_n)\le\mu(\psi)\) for every stochastic FO, where \(P_n=\|K_n\ket\psi\|^2\) and \(\ket{\psi_n}=K_n\ket\psi/\sqrt{P_n}\) for \(P_n>0\). Its convex-roof extension to mixed states is therefore a convex monotone under FO~\cite{Vidal2000}.
(b)~\(\Pst(\psi\to\phi)\le\mu(\psi)/\mu(\phi)\); for a balanced target, \(\Pst(\psi\to\Psi_\vphi)\le d\,\cN(\vphi)\mu(\psi)\).
(c)~\(\mu(\psi)\le1/[d\lmin(S)]\); we call states that attain this bound \(\mu\)-maximal. For equal overlaps the \(\mu\)-maximal states are \(\ket{\Psi_0}\) if \(s<0\) and the class-(A) states if \(s>0\).
\end{proposition}

\begin{proof}
(a) Write \(K_n=\sum_k\xi^{(n)}_k\ketbra{c_{h_n(k)}}{c_k^\perp}\). If \(h_n\) is not a permutation, some \(\ket{c_j}\) is absent from \(K_n\ket\psi\), and \(\mu(\psi_n)=0\). Otherwise \(K_n\ket\psi=\sum_k\xi^{(n)}_kx_k\ket{c_{h_n(k)}}\), so \(P_n\mu(\psi_n)\le|\xi^{(n)}_jx_j|^2=|\xi^{(n)}_j|^2\mu(\psi)\) for an index \(j\) with \(|x_j|^2=\mu(\psi)\). Summing over \(n\) and using \(\sum_n|\xi^{(n)}_j|^2=\sum_n\|K_n\ket{c_j}\|^2\le1\) proves the claim. The statement on the convex roof is the general construction of Ref.~\cite{Vidal2000}. (b) Every successful outcome is \(\ket\phi\), so (a) gives \(\Pst\mu(\phi)\le\mu(\psi)\). (c) Since \(1=x^\dagger Sx\ge\lmin(S)\|x\|^2\ge d\lmin(S)\mu(\psi)\), the bound holds, with equality iff \(x\) is a \(\lmin\)-eigenvector of \(S\) with equal moduli. For equal overlaps the \(\lmin\)-eigenspace is \({\rm span}\{\ones\}\) for \(s<0\) and \(\ones^\perp\) for \(s>0\); its balanced elements are \(\ket{\Psi_0}\) and the states with \(\zeta=0\), respectively.
\end{proof}

The \(\mu\)-maximal states are thus the balanced states that \(\cT\) leaves invariant with its smallest eigenvalue. For equal overlaps they are golden only for \(s\le0\) or \(d=2\), and Theorem~\ref{thm:price} quantifies what they lose otherwise. In the orthonormal limit the bound (b) for a balanced target is the optimal coherence-distillation probability \(d\min_j|x_j|^2\)~\cite{Fang2018}. For nonorthogonal reference states it is in general not attained; the next section gives the exact value.

%=====================================================================
\section{Optimal conversion probability}\label{sec:sdp}
%=====================================================================

\begin{figure*}[t]
\centering
\includegraphics[width=\textwidth]{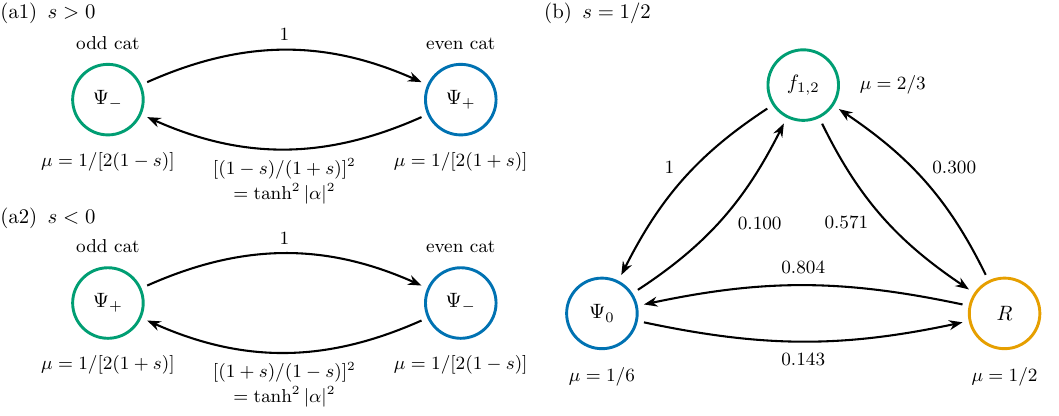}
\caption{Balanced states are inequivalent resources. Each arrow points from source to target and carries the optimal conversion probability \(\Pst\); \(\mu\) is the minimal weight, and the color marks the state with the largest (green) and the smallest (blue) minimal weight. (a) Qubit, (a1) for \(0<s<1\) and (a2) for \(-1<s<0\) (Corollary~\ref{cor:qubit}); the sign change \(\ket{c_1}\to-\ket{c_1}\) of Sec.~\ref{sec:equal} maps one row onto the other and changes only the labels \(\Psi_\pm\). In both rows the state with the larger minimal weight becomes the other one with certainty. For coherent states, \(\ket{c_0}=\ket\alpha\) and \(\ket{c_1}=\pm\ket{-\alpha}\) give the two rows, with \(s=\pm e^{-2|\alpha|^2}\): the odd cat state becomes even with certainty, and the reverse succeeds with probability \(\tanh^2|\alpha|^2\). (b) Qutrit, \(s=1/2\): the in-phase state \(\ket{\Psi_0}\), the Fourier states \(\ket{f_{1,2}}\) (related by a free permutation), and the real state \(\ket R=\ket{\Psi_{(0,\pi,\pi)}}\); values from the exact expressions of Table~\ref{tab:qutrit}. The Fourier states have the largest minimal weight, yet they yield \(\ket R\) only with probability \(4/7\approx0.571\).}
\label{fig:graphs}
\end{figure*}

\subsection{Semidefinite program}\label{sec:program}

The optimal probability~\eqref{eq:Pdef} is the value of a semidefinite program over the permutation branches of Lemma~\ref{lem:branches}(a).

\begin{theorem}[Optimal probability~\cite{Theurer2017}]
\label{thm:sdp}
For full-support \(x\) and \(t\) and any Gram matrix \(S\), the optimal probability \(\Pst(\psi\to\phi)\) is
\begin{align}
\Pst&=\max_{w\ge0}\Big\{\textstyle\sum_\pi w_\pi:\ \sum_\pi w_\pi\,\Pi_\pi^TG_\phi\Pi_\pi\le G_\psi\Big\}\label{eq:primal}\\
&=\min_{Y\ge0}\big\{\tr(YG_\psi):\,\tr(Y\Pi_\pi^TG_\phi\Pi_\pi)\ge1\,\forall\pi\big\}.\label{eq:dual}
\end{align}
For an optimal \(w\), the branches \(\sqrt{w_\pi}\,\cX_\phi U_\pi\cX_\psi^{-1}\), completed by Lemma~\ref{lem:branches}(b), form an optimal protocol.
\end{theorem}

Here \(w_\pi\) is the probability with which branch \(\pi\) succeeds. Theurer \emph{et al.} derived this program as an operator inequality for the branch operators~\cite{Theurer2017}; Eq.~\eqref{eq:primal} writes it with component Gram matrices, and its case \(\sum_\pi w_\pi=1\) is the Gram majorization criterion of Ref.~\cite{Torun2026b}. The constraint is a matrix inequality: a weighted sum of permuted copies of the target's \(G_\phi\) must fit beneath the source's \(G_\psi\). In coherence theory all these matrices are diagonal, and Eq.~\eqref{eq:primal} becomes a linear program over the populations. In RTS the off-diagonal entries \(\bar x_jS_{jk}x_k\) carry the overlaps and the relative phases, and different permutation branches probe different directions, so both the choice and the combination of branches matter. For Haar-random sources and the target \(\ket{f_1}\) at \(s=\tfrac12\), for example, the mean success probability is \(0.140\) for the identity branch alone, \(0.176\) for the best single branch, and \(0.180\) at the optimum for \(d=3\), and \(0.091\), \(0.131\), and \(0.143\) for \(d=4\) (Appendix~\ref{app:numerics}).

Every feasible \(w\) in Eq.~\eqref{eq:primal} is an explicit protocol, and every feasible \(Y\) in Eq.~\eqref{eq:dual} certifies an upper bound. The simplest certificates are test vectors: evaluating the constraint of Eq.~\eqref{eq:primal} in a single direction \(y\) gives
\begin{equation}
\Pst(\psi\to\phi)\le\frac{y^\dagger G_\psi y}{\min_\pi y^\dagger\Pi_\pi^TG_\phi\Pi_\pi y},
\label{eq:testvector}
\end{equation}
where, with \(\ket u=\sum_jy_j\ket{c_j}\) as in Eq.~\eqref{eq:Gform}, the numerator is the squared norm of the \emph{test state} \(\cX_\psi\ket u\) and the denominator is \(\min_\pi\|\cX_\phi U_\pi\ket u\|^2\). For a normalized test state the bound has a direct reading: a branch that succeeds on \(\ket\psi\) with probability \(w_\pi\) is triggered by the test state with probability \(w_\pi\|\cX_\phi U_\pi\ket u\|^2\), and these probabilities sum to at most one.

\subsection{Interference principle}\label{sec:principle}

For every pair of states the optimal probability is bounded from both sides, and it takes a closed form when the component Gram matrix of the source or of the target does not change under any relabeling of the components.

\begin{theorem}[Interference bounds and permutation invariance]
\label{thm:inphase}
Let \(x\) and \(t\) have full support, with reference states of Gram matrices \(S\) and \(S'\).
(a)~\(d\,\lmin(G_\psi)\min\{1,1/\tau(\phi)\}\le\Pst(\psi\to\phi)\le\min\{1,\tau(\psi)/\tau(\phi)\}\).
(b)~If \(\Pi_\pi^TG_\psi\Pi_\pi=G_\psi\) for all \(\pi\), then \(\Pst(\psi\to\phi)=\min\{1,\tau(\psi)/\tau(\phi)\}\), attained with equal branch probabilities over any doubly transitive group of permutations.
(c)~If \(\Pi_\pi^TG_\phi\Pi_\pi=G_\phi\) for all \(\pi\), then \(\Pst(\psi\to\phi)=\lmin(G_\psi,G_\phi)\), attained by the single filter \(\sqrt{\Pst}\,\cX_\phi\cX_\psi^{-1}\).
\end{theorem}

Here \(\lmin(G_1,G_2)\) is the smallest generalized eigenvalue of the pencil \((G_1,G_2)\), that is, the largest \(\lambda\) with \(\lambda G_2\le G_1\).

The upper bound of part (a) is an interference principle for full-support states: FO cannot make their components interfere more destructively with certainty, and they can raise \(\tau\) only at the cost of probability, \(\Pst\,\tau(\phi)\le\tau(\psi)\). It comes from the dual certificate \(Y\propto\one-J/d\) in Eq.~\eqref{eq:dual}, which shows more: \(\sum_nP_n\tau(\psi_n)\le\tau(\psi)\) for the full-support outcomes \(\ket{\psi_n}\) of every stochastic FO, so that \(\tau\), set to zero for states without full support, is a strong monotone like \(\mu\). In coherence theory \(\tau=1\) for every full-support state, and the bound is trivial. The restriction to full support matters, because a free state, which has a single component and \(\tau=1\), can be prepared from any state with certainty. The lower bound of part (a) is reached with equal probabilities on all permutation branches, a uniform \emph{twirl}. Its two factors are the optimal probabilities of the conversions \(\ket\psi\to\ket{\Phi_d}\) and \(\ket{\Phi_d}\to\ket\phi\), where \(\ket{\Phi_d}\) is a maximally coherent state of an orthonormal basis (Theorem~\ref{thm:bridge}), so it is the probability of the route through maximal coherence.

A component Gram matrix is permutation invariant iff all \(|x_j|\) are equal and all \(\bar x_jS_{jk}x_k\) with \(j\neq k\) are real and equal. This holds for the maximally coherent states, for the qubit states \(\ket{\Psi_\pm}\) when \(s\) is real, and, for equal overlaps \(s\neq0\) and \(d\ge3\), only for the in-phase state, with \(G_{\Psi_0}=S/(d\lambda_0)\). More generally, a permutation-invariant \(G_\psi\) forces \(S_{jk}=s\,e^{i(\varphi_j-\varphi_k)}\) for \(j\neq k\), with real \(s\) and \(x_j=|x_j|e^{i\varphi_j}\): up to the rephasing \(\ket{c_j}\to e^{i\varphi_j}\ket{c_j}\), every permutation-invariant source is the in-phase state of equal real overlaps. Part (b) shows that the exchange rate of part (a) is exact for these sources, for every target in any reference set. For equal overlaps, (b) and (c) give the in-phase state as source and as target,
\begin{align}
\Pst(\Psi_0\to\phi)&=\min\{1,\tau_0/\tau(\phi)\},\label{eq:from}\\
\Pst(\psi\to\Psi_0)&=d\lambda_0\lmin(G_\psi,S)=d\lambda_0\lmin(\cX_\psi^\dagger\cX_\psi),\label{eq:to}
\end{align}
where \(\tau_0=\tau(\Psi_0)\) is given by Eq.~\eqref{eq:range} and \(\lmin(\cX_\psi^\dagger\cX_\psi)\) is the squared smallest singular value of the coefficient operator.

Equation~\eqref{eq:from} depends on the target only through \(\tau(\phi)\), that is, through the squared Euclidean norm \(\|t\|^2\) of its coefficients. This has two consequences. First, for \(s\le0\), \(\tau_0\) is the largest value of \(\tau\), and Eq.~\eqref{eq:from} equals one for every full-support target, in line with the fact that \(\ket{\Psi_0}\) is then a universal source, that is, a golden state~\cite{Senyasa2022}; Eq.~\eqref{eq:to} is then the optimal probability of distilling this universal source and hence a lower bound on every \(\Pst(\psi\to\phi)\). Second, for \(s>0\) it is below one for every target other than \(\ket{\Psi_0}\); for balanced targets, which have \(\|t\|^2=1/\cN(\vphi)\),
\begin{equation}
\Pst(\Psi_0\to\Psi_\vphi)=\frac{(d-1)\lambda_1\,\cN(\vphi)}{\lambda_0\,[d-\cN(\vphi)]},
\label{eq:fromBSS}
\end{equation}
which is smallest on class (A), where it equals \(\tau_0/\tau_A\). Conversely, by Eq.~\eqref{eq:to}, every class-(A) state yields \(\ket{\Psi_0}\) with certainty for \(s>0\) and with probability \((\lambda_0/\lambda_1)^2\) for \(s<0\).

In Eq.~\eqref{eq:to} the squared smallest singular value of the coefficient operator replaces the minimal weight of coherence theory. Expanding it in the overlap shows how nonorthogonality enters. For a fixed direction of \(x\) whose minimal modulus is attained at a single index \(j^\ast\),
\begin{equation}
\frac{\Pst(\psi\to\Psi_0)}{d\lambda_0\,\mu(\psi)}=1-s^2\sum_{k\neq j^\ast}\frac{|x_k-x_{j^\ast}|^2}{|x_k|^2-|x_{j^\ast}|^2}+O(s^3).
\label{eq:secondorder}
\end{equation}
The minimal-weight bound is thus exact to first order. The penalty is quadratic in \(s\), and it depends on the relative \emph{phases} of the coefficients, not only on their moduli.

\subsection{Qubits and cat states}\label{sec:cats}

For qubits with real \(s\), the two real balanced states \(\ket{\Psi_\pm}\) have permutation-invariant component Gram matrices, so Theorem~\ref{thm:inphase}(c) gives \(\Pst\) for every source.

\begin{corollary}[Qubits]
\label{cor:qubit}
For \(d=2\), real \(s\), and \(\ket\psi=x_0\ket{c_0}+x_1\ket{c_1}\),
\begin{equation}
\Pst(\psi\to\Psi_\pm)=\frac{1\pm s}{1-s^2}\Big(Q_\pm-\sqrt{Q_\pm^2-\Delta}\Big),
\label{eq:qubit}
\end{equation}
with \(Q_\pm=|x_0|^2+|x_1|^2\mp2s^2{\rm Re}(\bar x_0x_1)\) and \(\Delta=4(1-s^2)^2|x_0x_1|^2\). In particular, \(\Pst(\Psi_-\to\Psi_+)=1\) and \(\Pst(\Psi_+\to\Psi_-)=[(1-s)/(1+s)]^2\) for \(s>0\), whereas \(\Pst(\Psi_+\to\Psi_-)=1\) and \(\Pst(\Psi_-\to\Psi_+)=[(1+s)/(1-s)]^2\) for \(s<0\).
\end{corollary}

The two cases express a single fact. The stronger of the two states \(\ket{\Psi_\pm}\) is the one whose components interfere destructively, that is, whose coefficient vector is the eigenvector of \(S\) with the smaller eigenvalue \(1-|s|\); it is the \(\mu\)-maximal and golden state of the qubit, becomes the other state with certainty, and is reached from it at best with probability \([(1-|s|)/(1+|s|)]^2<1\). The states \(\ket{\Psi_\pm}\) are thus inequivalent resources for every \(s\neq0\), and the sign of \(s\) only decides whether the stronger one is labeled \(\ket{\Psi_-}\) or \(\ket{\Psi_+}\).

\begin{figure}[t]
\centering
\includegraphics[width=\columnwidth]{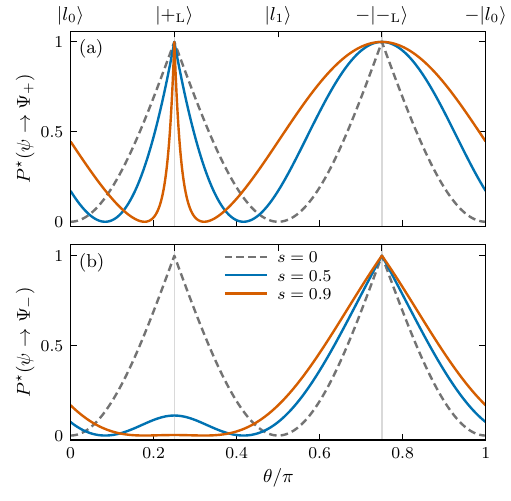}
\caption{Optimal qubit distillation, Eq.~\eqref{eq:qubit}, for the real sources \(\ket\psi=\cos\theta\ket{l_0}+\sin\theta\ket{l_1}\) and \(s=0\) (coherence theory), \(0.5\), and \(0.9\). (a) Target \(\ket{\Psi_+}=\ket{+_{\rm L}}\), the even cat state for the coherent states \(\ket{c_{0,1}}=\ket{\pm\alpha}\). (b) Target \(\ket{\Psi_-}=\ket{-_{\rm L}}\), the odd cat state. Gray vertical lines mark the sources \(\ket{\pm_{\rm L}}=\ket{\Psi_\pm}\). The zeros occur at the free states \(\ket\psi\propto\ket{c_j}\). For \(s<0\) the panels exchange their roles: each shows the other panel for \(|s|\), reflected about \(\theta=\pi/2\).}
\label{fig:qubit}
\end{figure}

Figure~\ref{fig:qubit} shows Eq.~\eqref{eq:qubit} for real sources. Even the L\"owdin-frame states \(\ket{l_j}\), which are free in the coherence theory of the L\"owdin frame, carry superposition: for \(s=\tfrac12\), \(\Pst(l_0\to\Psi_+)=3-2\sqrt2\approx0.17\). The two panels differ because superposition coefficients are L\"owdin coefficients transformed by \(S^{-1/2}\), the inverse of the ellipse map in Fig.~\ref{fig:qubit_geometry}(b): for \(s>0\), a small deviation from \(\ket{+_{\rm L}}\) is magnified in the coefficients by \(\sqrt{\lambda_0/\lambda_1}\), and a deviation from \(\ket{-_{\rm L}}\) is reduced by the same factor.

For the coherent states \(\ket{c_0}=\ket\alpha\) and \(\ket{c_1}=\ket{-\alpha}\), \(s=e^{-2|\alpha|^2}>0\) and \((1-s)/(1+s)=\tanh|\alpha|^2\) [Fig.~\ref{fig:graphs}(a1)], so the stronger state is the odd cat state; the convention \(\ket{c_1}=-\ket{-\alpha}\) gives \(s<0\) and labels it \(\ket{\Psi_+}\) [Fig.~\ref{fig:graphs}(a2)] without changing any probability. The optimal operation from the odd to the even cat state is
\begin{equation}
\begin{aligned}
K&=\ketbra{+_{\rm L}}{-_{\rm L}}+\frac{\lambda_1}{\lambda_0}\ketbra{-_{\rm L}}{+_{\rm L}}\\
&=\sqrt{\frac{\lambda_1}{\lambda_0}}\big(\ketbra{c_0}{c_0^\perp}-\ketbra{c_1}{c_1^\perp}\big),
\end{aligned}
\label{eq:qubitK}
\end{equation}
completed by \(\sqrt{1-(\lambda_1/\lambda_0)^2}\ketbra{c_0}{+_{\rm L}}\). It maps \(\ket{\Psi_-}\) to \(\ket{\Psi_+}\) with certainty and \(\ket{\Psi_+}\) to \(\ket{\Psi_-}\) with the optimal probability \((\lambda_1/\lambda_0)^2\). The asymmetry comes from the ellipse of Fig.~\ref{fig:qubit_geometry}(b): the phase flip \(\ketbra{c_0}{c_0^\perp}-\ketbra{c_1}{c_1^\perp}\), which has the superposition-free form~\eqref{eq:FOkraus} but is not unitary, exchanges the long and short axes of the ellipse. For \(s<0\) the optimal operation is the same phase flip normalized to unit operator norm, \(\sqrt{\lambda_0/\lambda_1}\,(\ketbra{c_0}{c_0^\perp}-\ketbra{c_1}{c_1^\perp})\), and it maps \(\ket{\Psi_+}\) to \(\ket{\Psi_-}\) with certainty. For \(\ket{c_{0,1}}=\ket{\pm\alpha}\), the eigenvalue equation \(\hat a\ket{\pm\alpha}=\pm\alpha\ket{\pm\alpha}\) shows that the photon annihilation operator acts on their span as
\begin{equation}
\hat a=\alpha\big(\ketbra{c_0}{c_0^\perp}-\ketbra{c_1}{c_1^\perp}\big).
\label{eq:annihilation}
\end{equation}

The optimal operation~\eqref{eq:qubitK} is therefore a normalized photon subtraction: the annihilation operator on the span of \(\ket{\pm\alpha}\), rescaled to unit operator norm. The heralded photon subtraction with a weakly reflecting beam splitter that generates small optical cat states from squeezed light~\cite{Dakna1997,Ourjoumtsev2006} applies the same operator with a small prefactor.

Because the odd and even cat states and the maximally coherent qubit \(\ket{\Phi_2}\) of an orthonormal basis have permutation-invariant component Gram matrices, Theorem~\ref{thm:inphase}(b) gives every conversion among them, also between different amplitudes, that is, by superposition-free maps from the reference states \(\ket{\pm\alpha}\) to \(\ket{\pm\beta}\). With \(\tau(\Phi_2)=1\) and
\begin{equation}
\tau({\rm odd}_\alpha)=\coth|\alpha|^2,\quad\tau({\rm even}_\alpha)=\tanh|\alpha|^2,
\label{eq:taucats}
\end{equation}
every optimal probability is \(\min\{1,\tau(\psi)/\tau(\phi)\}\), a product or ratio of hyperbolic tangents of \(|\alpha|^2\) and \(|\beta|^2\) (Table~\ref{tab:cats}).

\begin{table}[t]
\caption{Optimal conversion probabilities among the odd and even cat states built from \(\ket{\pm\alpha}\) (rows, source) and from \(\ket{\pm\beta}\) (columns, target), and the maximally coherent qubit \(\ket{\Phi_2}\), with \(T_\alpha=\tanh|\alpha|^2\) and \(T_\beta=\tanh|\beta|^2\). Every entry is \(\min\{1,\tau(\psi)/\tau(\phi)\}\) [Theorem~\ref{thm:inphase}(b)].}
\label{tab:cats}
\begin{ruledtabular}
\begin{tabular}{lccc}
 & \({\rm odd}_\beta\) & \({\rm even}_\beta\) & \(\Phi_2\)\\\hline
\({\rm odd}_\alpha\) & \(\min\{1,T_\beta/T_\alpha\}\) & \(1\) & \(1\)\\
\({\rm even}_\alpha\) & \(T_\alpha T_\beta\) & \(\min\{1,T_\alpha/T_\beta\}\) & \(T_\alpha\)\\
\(\Phi_2\) & \(T_\beta\) & \(1\) & \(1\)\\
\end{tabular}
\end{ruledtabular}
\end{table}

The values~\eqref{eq:taucats} are the mean photon numbers of the two cat states in units of \(|\alpha|^2\): destructive interference raises the mean photon number above that of the components, and the optimal probability \(\tanh^2|\alpha|^2\) of the conversion from the even to the odd cat state is the ratio of the mean photon numbers, as expected for a photon subtraction, whose success probability is proportional to the mean photon number. Odd cat states can thus be enlarged with certainty but reduced only probabilistically, and even cat states reduced with certainty but enlarged only probabilistically (Table~\ref{tab:cats}); the enlargement of optical cat states is an experimental goal in its own right~\cite{Lund2004,Sychev2017}. The ordering agrees with the limit \(\alpha\to0\), in which the even cat state approaches the vacuum, a coherent state, whereas the odd cat state approaches the one-photon state.

\subsection{Several copies}\label{sec:copies}

Collective operations on several copies can do better than converting the copies one at a time. For \(N\) copies the reference states are the \(2^N\) products \(\ket{c_{j_1}}\otimes\dots\otimes\ket{c_{j_N}}\), with Gram matrix \(S^{\otimes N}\), and Theorem~\ref{thm:sdp} applies with \(G_{\psi^{\otimes N}}=G_\psi^{\otimes N}\) and with all \((2^N)!\) permutations of the product reference states.

\begin{proposition}[Copies]
\label{prop:copies}
For \(d=2\), \(0<s<1\), and \(r=(1-s)/(1+s)\),
\begin{equation}
r^{2N}\le\Pst\big(\Psi_+^{\otimes N}\to\Psi_-^{\otimes N}\big)\le r^{\,N+1},
\label{eq:copies}
\end{equation}
with equality on the right for \(N\le3\).
\end{proposition}

The left-hand side is the probability of converting the copies one at a time. For the cat states built from \(\ket{\pm\alpha}\), \(r=\tanh|\alpha|^2\): two even cat states become two odd cat states with probability \(\tanh^3|\alpha|^2\), and three with probability \(\tanh^4|\alpha|^2\), so the second and the third copy each cost a single factor \(\tanh|\alpha|^2\) instead of two. The optimal protocols for \(N=2\) and \(N=3\) are single branches whose permutation is a linear map of the index strings \((j_1,\dots,j_N)\in\mathbb Z_2^N\); for \(N=2\) it is \((j_1,j_2)\mapsto(j_1,j_1\oplus j_2)\), which exchanges the product reference states \(\ket{c_1}\ket{c_0}\) and \(\ket{c_1}\ket{c_1}\). For \(N=4\) no single branch of this kind attains the upper bound, but mixtures of such branches attain it for \(r\ge0.225\) (Appendix~\ref{app:numerics}); whether the bound is attained for smaller \(r\) and for \(N\ge5\) is open.

For conversions to and from coherence, Theorem~\ref{thm:inphase} solves the problem for every number of copies, because \(\ket{\Phi_d}^{\otimes N}\) is the maximally coherent state of the product basis and has a permutation-invariant component Gram matrix.

\begin{corollary}[Copies and coherence]
\label{cor:copies}
For any \(S\), full-support \(x\) and \(t\), and every \(N\),
\begin{equation}
\begin{aligned}
\Pst\big(\psi^{\otimes N}\to\Phi_d^{\otimes N}\big)&=\big[d\,\lmin(G_\psi)\big]^N,\\
\Pst\big(\Phi_d^{\otimes N}\to\phi^{\otimes N}\big)&=\min\Big\{1,\frac{d^N-1}{d^N\|t\|^{2N}-1}\Big\}.
\end{aligned}
\label{eq:copiescoh}
\end{equation}
\end{corollary}

The distillation of coherence from superposition is thus multiplicative: collective operations do not help. The reverse conversion profits from them whenever it is probabilistic for a single copy, \(\tau(\phi)>1\). For the odd cat state, \(\|t\|^2=1/(1-s)\), and the probability per copy tends to \(1-s=1-e^{-2|\alpha|^2}\) for large \(N\), instead of \(\tanh|\alpha|^2\) for a single copy.

\subsection{Two reference states}\label{sec:two}

Two reference states can be treated completely. For \(d=2\) the program~\eqref{eq:primal} has only two branches, the identity and the exchange of the reference states. For \(2\times2\) matrices \(H_1\) and \(H_2\) let \(\Omega(H_1,H_2)=\tr H_1\,\tr H_2-\tr(H_1H_2)\), so that \(\det(H_1-wH_2)=\det H_1-w\,\Omega(H_1,H_2)+w^2\det H_2\). Let \(G_{\rm ex}=\Pi G_\phi\Pi\), with \(\Pi\) the permutation matrix of the exchange, be the component Gram matrix of the target with its two components exchanged, and let \(\Omega_{\rm id}=\Omega(G_\psi,G_\phi)\), \(\Omega_{\rm ex}=\Omega(G_\psi,G_{\rm ex})\), and \(\Omega_\pm=\Omega(G_\phi,G_{\rm ex})\pm2\det G_\phi\).

\begin{proposition}[Two reference states]
\label{prop:qubit}
Let \(d=2\), let \(s\) be real, and let \(x\) and \(t\) have full support. The identity branch alone and the exchange branch alone succeed with the probabilities
\begin{equation}
P_{\rm id,ex}=\frac{\Omega_{\rm id,ex}-\sqrt{\Omega_{\rm id,ex}^2-4\det G_\psi\det G_\phi}}{2\det G_\phi},
\label{eq:Pid}
\end{equation}
and we set
\begin{equation}
\begin{aligned}
P_{\rm mix}&=\frac{\Omega_{\rm id}+\Omega_{\rm ex}-\sqrt{\Omega_0}}{\Omega_+},\\
\Omega_0&=(\Omega_{\rm id}+\Omega_{\rm ex})^2-4\Omega_+\det G_\psi\\
&\quad-(\Omega_+/\Omega_-)\,(\Omega_{\rm id}-\Omega_{\rm ex})^2 .
\end{aligned}
\label{eq:Pmix}
\end{equation}
If \(\Omega_->0\) and \(|\Omega_{\rm id}-\Omega_{\rm ex}|\le\Omega_-P_{\rm mix}\), then \(\Pst(\psi\to\phi)=P_{\rm mix}\), attained with the branch probabilities \(w_{\rm id,ex}=\frac12P_{\rm mix}\pm(\Omega_{\rm ex}-\Omega_{\rm id})/(2\Omega_-)\). Otherwise \(\Pst(\psi\to\phi)=\max\{P_{\rm id},P_{\rm ex}\}\).
\end{proposition}

Here \(\Omega_-=(|t_0|^2-|t_1|^2)^2+4s^2[{\rm Im}(\bar t_0t_1)]^2\), which for \(s\neq0\) vanishes iff \(t_1=\pm t_0\), the targets of Corollary~\ref{cor:qubit}, for which the two branches coincide. For the coherent states \(\ket{c_0}=\ket\alpha\) and \(\ket{c_1}=\ket{-\alpha}\), the proposition gives every conversion among their superpositions. The balanced ones are the cat states with a relative phase, \(\ket{\Psi_\varphi}=(\ket{c_0}+e^{i\varphi}\ket{c_1})/\sqrt{2\cN(\varphi)}\) with \(\cN(\varphi)=1+s\cos\varphi\): the even cat state (\(\varphi=0\)), the odd cat state (\(\varphi=\pi\)), and the Yurke--Stoler cat state (\(\varphi=\pi/2\)), which a Kerr nonlinearity generates from a coherent state~\cite{Yurke1986}. They have \(\tau(\Psi_\varphi)=(1-s\cos\varphi)/(1+s\cos\varphi)\). The exchange of the reference states, a free unitary, maps \(\varphi\to-\varphi\) up to a global phase, so it suffices to take \(0\le\varphi\le\pi\). We denote the phase of a balanced target by \(\vartheta\).

\begin{corollary}[Cat states with a relative phase]
\label{cor:phase}
Let \(d=2\), let \(s\neq0\) be real, let \(0\le\varphi,\vartheta\le\pi\), and let \(P_\tau=\min\{1,\tau(\Psi_\varphi)/\tau(\Psi_\vartheta)\}\). If
\begin{equation}
\frac{\sin\varphi}{\cN(\varphi)}\le P_\tau\,\frac{\sin\vartheta}{\cN(\vartheta)},
\label{eq:phasecond}
\end{equation}
then \(\Pst(\Psi_\varphi\to\Psi_\vartheta)=P_\tau\). Otherwise, with \(\Upsilon=1-s^2\cos(\varphi-\vartheta)\),
\begin{equation}
\Pst(\Psi_\varphi\to\Psi_\vartheta)=\frac{\cN(\vartheta)}{\cN(\varphi)}\,\frac{\Upsilon-\sqrt{\Upsilon^2-(1-s^2)^2}}{1-s^2}<P_\tau.
\label{eq:phasesingle}
\end{equation}
\end{corollary}

Two quantities thus govern the cat states with a relative phase: the real part of the interference term of the two components, measured by \(\tau\), and its imaginary part, measured by \(\sin\varphi/\cN(\varphi)\). The interference bound of Theorem~\ref{thm:inphase}(a) is attained iff the target carries enough of the second, because the exchange branch reverses the sign of the imaginary part, and mixing the two branches can only reduce it. The odd cat state becomes every \(\ket{\Psi_\vartheta}\) with certainty, and the even cat state becomes \(\ket{\Psi_\vartheta}\) with probability \(\tau(\Psi_0)/\tau(\Psi_\vartheta)\), in agreement with Theorem~\ref{thm:inphase}(b). The Yurke--Stoler cat state is obtained with certainty from the odd cat state and with probability \(\tanh|\alpha|^2\) from the even cat state, but it becomes the even cat state only with probability \([1-s\sqrt{2-s^2}]/(1-s)<1\), although its components interfere more destructively, \(\tau(\Psi_{\pi/2})=1>\tau(\Psi_0)\). The even and the Yurke--Stoler cat states are therefore incomparable: neither becomes the other with certainty, and \(\tau\) alone does not order even the balanced states of two reference states.

\subsection{Optimal protocols}\label{sec:protocols}

The optimal protocols of Theorem~\ref{thm:inphase} have a simple form. For a permutation-invariant target, the \emph{L\"owdin filter}, the single branch \(\cX_\phi\cX_\psi^{-1}\) normalized to unit operator norm, is optimal. In the L\"owdin frame it is proportional to \(S^{1/2}DS^{-1/2}\) with \(D=\diag(t_j/x_j)\): a diagonal filter on the superposition coefficients, placed between the forward and the inverse L\"owdin maps. For a permutation-invariant source, the optimal twirl uses the branches of a doubly transitive group \(\mathcal G\) with equal probabilities,
\begin{equation}
K_g=\sqrt{\Pst/|\mathcal G|}\;\cX_\phi U_g\cX_\psi^{-1},\quad g\in\mathcal G;
\label{eq:twirl}
\end{equation}
for \(\ket{\Psi_0}\), \(\cX_{\Psi_0}\propto\one\), and the twirl is a random free permutation followed by the coefficient operator of the target. For prime-power \(d\) the affine group of the field with \(d\) elements requires only \(d(d-1)\) branches~\cite{DixonMortimer1996}.

%=====================================================================
\section{Inequivalence of balanced states}\label{sec:nogo}
%=====================================================================

\subsection{No universal source}\label{sec:nouniversal}

By Proposition~\ref{prop:mu}(b), a good source must have a large minimal weight. For equal overlaps with \(s\le0\) the in-phase state is the best source and reaches every state with certainty~\cite{Senyasa2022}. For \(s>0\) the candidates are the \(\mu\)-maximal class-(A) states, and one may expect them to play the role of the maximally coherent states. For \(d=2\) they do, by the sign change of Sec.~\ref{sec:equal}; for \(d\ge3\) they do not. For a class-(A) source \(\ket{\Psi_\vphi}\) and a target with coefficients \(t\), let
\begin{equation}
a_\pi=\Big|\sum_jt_{\pi(j)}\,e^{-i\varphi_j}\Big|^2
\label{eq:align}
\end{equation}
be the \emph{alignment} of the target with the source in branch \(\pi\), with maximum \(a_{\max}\) and mean \(\langle a\rangle=[\tau(\phi)-\lambda_1\|t\|^2]/s\) over the \(d!\) permutations.

\begin{theorem}[No universal balanced source]
\label{thm:nogo}
Let the overlaps be equal with \(0<s<1\), and let \(d\ge3\).
(i)~For every class-(A) state \(\ket{\Psi_\vphi}\) and every full-support target \(\ket\phi\),
\begin{equation}
\Pst(\Psi_\vphi\to\phi)\le\frac{d-2}{(d-2)\,\tau(\phi)-s\,(a_{\max}-\langle a\rangle)}.
\label{eq:witness}
\end{equation}
(ii)~For \(d=3\) and every full-support target, with the populations \(\rho_k=|\braket{f_k}{\phi}|^2\) of the target in the Fourier basis and \(\rho_<=\min\{\rho_1,\rho_2\}\),
\begin{equation}
\Pst(f_{1,2}\to\phi)=\min\Big\{1,\,\Big[1+3s\Big(\frac{\rho_<}{\lambda_1}-\frac{\rho_0}{\lambda_0}\Big)\Big]^{-1}\Big\}.
\label{eq:qutritall}
\end{equation}
In particular, \(\Pst(f_{1,2}\to R)=1/\tau(R)=(3-2s)/(3+s)\).
(iii)~No pure state yields every balanced state with certainty. In particular, no golden state exists under FO.
\end{theorem}

Part (i) follows from the trace of the constraint~\eqref{eq:primal} against the projector orthogonal to \(\ones\) and to the phase vector of the source, a mixed-state version of the test vector~\eqref{eq:testvector}. For a balanced target \(\ket{\Psi_\vth}\) it reads
\begin{equation}
\Pst(\Psi_\vphi\to\Psi_\vth)\le\frac{(d-2)\,\cN(\vth)}{d-1+s-\cN(\vth)-sd\max_\pi|\zeta_\pi|^2},
\label{eq:witnessBSS}
\end{equation}
where \(\zeta_\pi=\frac1d\sum_je^{i(\vartheta_{\pi(j)}-\varphi_j)}\). For a class-(A) target, \(\cN(\vth)=\lambda_1\), and the bound is below one unless \(\vth\) is a permutation of \(\vphi\) up to a global phase: a \(\mu\)-maximal source cannot reach most balanced targets with certainty.

Part (iii) follows from part (i) and Proposition~\ref{prop:mu}. A state that is not \(\mu\)-maximal yields no class-(A) target with certainty, because its minimal weight is smaller [Proposition~\ref{prop:mu}(b)]. A \(\mu\)-maximal state is a class-(A) state \(\ket{\Psi_\vphi}\) [Proposition~\ref{prop:mu}(c)], and by Eq.~\eqref{eq:witnessBSS} it fails for \(\ket R\) if \(d=3\), and for \(d\ge4\) for every class-(A) target whose pattern is not obtained from \(\vphi\) by a permutation and a global phase; such targets exist because class (A) is then a continuum. Mixed states need not be considered: if a channel maps a mixed state to a pure target, it maps every pure state in the range of that mixed state to the same target. Golden states are understood with respect to FO; under the maximal superposition-free operations they exist for every Gram matrix~\cite{Torun2026b}. That no golden state exists under FO for \(d\ge3\) and \(0<s<1\) was shown in Ref.~\cite{Torun2026b} by Gram majorization with a target of unequal moduli; for \(d=3\) and \(s=\tfrac12\), Ref.~\cite{Theurer2017} had excluded the candidates \(\ket{f_{1,2}}\) by a dual bound. Part (iii) strengthens this result, because the balanced targets alone already exclude a golden state, and parts (i) and (ii) add explicit failing targets, quantitative bounds, and, for the qutrit, the optimal probability for every full-support target.

\begin{table}[t]
\caption{Optimal conversion probabilities among the qutrit balanced states \(\ket{\Psi_0}\), \(\ket{f_1}\), and \(\ket R\) for \(s>0\) (\(\lambda_0=1+2s\), \(\lambda_1=1-s\)). All entries are exact; the last column is for \(s=\tfrac12\). The conversion \(f_1\to\Psi_0\) is certain, and \(\ket{f_2}\) behaves as \(\ket{f_1}\).}
\label{tab:qutrit}
\begin{ruledtabular}
\begin{tabular}{lcc}
conversion & \(\Pst\) & \(s=\tfrac12\)\\\hline
\(\Psi_0\to f_1\) & \(\dfrac{2\lambda_1^2}{\lambda_0(2+s)}\) & \(0.100\)\\[8pt]
\(\Psi_0\to R\) & \(\dfrac{\lambda_1(3-2s)}{\lambda_0(3+s)}\) & \(0.143\)\\[8pt]
\(f_1\to R\) & \(\dfrac{3-2s}{3+s}\) & \(0.571\)\\[8pt]
\(R\to\Psi_0\) & \(\dfrac{3(\sqrt{1+s}-\sqrt2\,s)^2}{\lambda_1(3-2s)}\) & \(0.804\)\\[8pt]
\(R\to f_1\) & \(\dfrac{6\lambda_1^2}{(2+s)(3-2s)}\) & \(0.300\)\\
\end{tabular}
\end{ruledtabular}
\end{table}

Part (ii) solves the qutrit, where class (A) consists of \(\ket{f_1}\) and \(\ket{f_2}\): the bound of part (i), capped at one, is attained for every target. Only two numbers enter, the in-phase population \(\rho_0=|\braket{\Psi_0}{\phi}|^2\) of the target and the smaller of its populations in the two Fourier states of opposite handedness. A Fourier state reaches both handednesses at no cost, since \(\ket{f_1}\) and \(\ket{f_2}\) are related by a free permutation; what costs probability is to populate both at once beyond what the in-phase population compensates, and the conversion is certain iff \(\lambda_0\,\rho_<\le\lambda_1\,\rho_0\). In the same terms \(\tau(\phi)=1+3s\,[(\rho_1+\rho_2)/(2\lambda_1)-\rho_0/\lambda_0]\), so for every achiral target (\(\rho_1=\rho_2\)), in particular for every target with real coefficients, \(\Pst(f_{1,2}\to\phi)=\min\{1,1/\tau(\phi)\}\), the optimal probability from a maximally coherent state [Theorem~\ref{thm:bridge}(ii)]: a Fourier state profits from its handedness only for chiral targets. The real state \(\ket R\) is achiral. For achiral targets the optimum is attained by the three cyclic branches alone, for \(\ket{f_1}\to\ket R\) with the Kraus operators
\begin{equation}
K_\ell=\sqrt{\tfrac{1-s}{3+s}}\;\sum_j\varepsilon_{j+\ell}\,\omega^{-j}\ketbra{c_{j+\ell}}{c_j^\perp},\quad\ell=0,1,2,
\label{eq:K1R}
\end{equation}
where \(\varepsilon=(1,-1,-1)\) and indices are taken modulo 3, and the upper bound is the test-vector bound~\eqref{eq:testvector} with the test state \(\ket{f_2}\), which triggers every branch with \(\tau(R)\) times the probability with which that branch succeeds. The proof rests on a reduction that holds more generally: for sources that are Fourier states of reference states with a circulant Gram matrix, such as the multicomponent cat states built from the coherent states \(\ket{\alpha\omega^j}\)~\cite{Grimsmo2020}, the program~\eqref{eq:primal} reduces to a linear program with one constraint per Fourier state (Lemma~\ref{lem:fourier} in Appendix~\ref{app:nogo}).

Table~\ref{tab:qutrit} and Fig.~\ref{fig:graphs}(b) give all conversions among the three distinguished qutrit states. The minimal weight orders them as \(\mu(\Psi_0)<\mu(R)<\mu(f_{1,2})\), but the conversions do not follow this order: \(\ket{f_{1,2}}\) yield \(\ket{\Psi_0}\) with certainty and \(\ket R\) only with probability \(4/7\) at \(s=\tfrac12\). The optimal Kraus operators for the two conversions out of \(\ket{\Psi_0}\) and \(\ket{R}\) into \(\ket{f_1}\) are
\begin{align}
\Psi_0\to f_1:&\quad K_\pi=\sqrt{\tfrac{\lambda_1}{3(2+s)}}\;Z_\omega U_\pi,\label{eq:K01}\\
R\to f_1:&\quad K_\pi=\sqrt{\Pst/6}\;\cX_{f_1}U_\pi\cX_R^{-1},\label{eq:KR1}
\end{align}
with \(\pi\in\Sym_3\) and \(Z_\omega=\sum_k\omega^k\ketbra{c_k}{c_k^\perp}\). Protocol~\eqref{eq:K01} is a random free permutation followed by the phase filter \(Z_\omega\).

\subsection{Price of frustration}\label{sec:price}

The failure of the \(\mu\)-maximal states has a transparent origin. The component Gram matrix of a class-(A) state is
\begin{equation}
G_\psi=\frac1d\Big(\one+\frac{s}{\lambda_1}\,vv^\dagger\Big),\quad v=\bar e_\vphi=(e^{-i\varphi_j})_j\perp\ones,
\label{eq:GclassA}
\end{equation}
a multiple of the identity plus a rank-one excess along the source's own phase vector. Orthogonal to \(v\) it coincides with the component Gram matrix \(\one/d\) of a maximally coherent state, and a target profits from the excess only to the extent that its permuted component Gram matrices are aligned with \(v\), which is what the alignment~\eqref{eq:align} measures. The class-(A) states thus fail as universal sources of superposition. The next theorem shows how close they come.

\begin{theorem}[Price of frustration]
\label{thm:price}
Let the overlaps be equal with \(0<s<1\), let \(d\ge3\), and let \(\ket\psi\) be a class-(A) state.
(i)~For every full-support target,
\begin{equation}
\begin{aligned}
\Pst(\psi\to\phi)&\ge\min\{1,1/\tau(\phi)\}\\
&=\Pst(\Phi_d\to\phi)\ge1/\tau_A,
\end{aligned}
\label{eq:pricebound}
\end{equation}
where \(\ket{\Phi_d}\) is a maximally coherent state of an orthonormal basis.
(ii)~The first inequality is an equality whenever \(a_\pi\) is the same for all \(\pi\), in particular for every target in which all coefficients but one are equal. Among these targets,
\begin{equation}
\ket{\phi_k}\propto(d-1)\ket{c_k}-\sum_{j\neq k}\ket{c_j}
\label{eq:hardest}
\end{equation}
has \(\tau(\phi_k)=\tau_A\). Hence the minimum of \(\Pst(\psi\to\phi)\) over all full-support targets is \(1/\tau_A\), for every class-(A) state.
\end{theorem}

Part (i) is the lower bound of Theorem~\ref{thm:inphase}(a), because \(d\,\lmin(G_\psi)=1\) by Eq.~\eqref{eq:GclassA}: target by target, the class-(A) states are at least as good sources as maximal coherence itself. Part (ii) follows from Theorem~\ref{thm:nogo}(i). It shows that they do no better in the worst case, and it identifies the obstruction. The target~\eqref{eq:hardest} is the reference state \(\ket{c_k}\) with its in-phase part removed, \(\ket{\phi_k}\propto(\one-\ketbra{\Psi_0}{\Psi_0})\ket{c_k}\). It has the same, largest, value \(\tau_A\) as the source, so the components of both interfere equally destructively in total; but whereas the source shares this interference evenly among all pairs of components, the target puts all of it between one heavy component and the \(d-1\) light ones, which interfere constructively with one another. In every branch such a target is equally aligned with the source, so the phase pattern of the source gives no advantage over a maximally coherent state. Interference cannot be redistributed among the components with certainty, and \(\tau\) alone does not decide convertibility. The hardest target is not unique: for sources with \(\sum_je^{2i\varphi_j}=0\), which include \(\ket{f_1}\) for every \(d\ge3\), the value \(1/\tau_A\) is also approached for targets that tend to \((\ket{c_j}-\ket{c_k})/\sqrt{2\lambda_1}\), \(j\neq k\), and so concentrate the interference in a single pair of components (Appendix~\ref{app:nogo}). For \(d=3\) the worst case \(1/\tau_A=2(1-s)/(2+s)\) lies below the value \((3-2s)/(3+s)\) for the balanced target \(\ket R\).

Combined with Eq.~\eqref{eq:from}, the worst-case probability of the \(\mu\)-maximal states is \(1\) for \(s\le0\), where the in-phase state is golden, and \({1/\tau_A=(d-1)(1-s)/(d-1+s)}\) for \(0<s<1\). It is continuous at \(s=0\) but has a kink there, with slope \(-d/(d-1)\) on the positive side: orthonormal reference states sit at a nonanalytic point between unfrustrated and frustrated interference, which the lifted trine states cross, and in the worst case a small positive overlap costs probability to first order. For \(d=2\) there is no price, because \(\ket{\Psi_-}\) is golden, and for \(d\to\infty\) the worst case tends to \(1-s\). Numerically, the \(\mu\)-maximal states, the best sources of coherence, are not quite the best universal sources: for \(d=3\) and \(s=\tfrac12\), a global search finds no target below \(0.406\) for a small unbalanced perturbation of \(\ket{f_1}\), whereas \(1/\tau_A=0.4\).

\subsection{L\"owdin maps between coherence and superposition}\label{sec:lowdinmaps}

Although they are not universal sources of superposition, the class-(A) states are, for \(0<s<1\), exactly the deterministic sources of coherence. The L\"owdin maps connect the two theories directly. The conversions of the next theorem are by superposition-free maps between reference sets (Sec.~\ref{sec:FO}), with an orthonormal basis \(\{\ket j\}\) on one side: from \(\{\ket{c_j}\}\) to \(\{\ket j\}\) in part (i), from \(\{\ket j\}\) to \(\{\ket{c_j}\}\) in part (ii), and between two sets of nonorthogonal reference states in part (iii).

\begin{theorem}[L\"owdin maps]
\label{thm:bridge}
(i)~\emph{Superposition \(\to\) coherence.} For any \(S\), the optimal probability of obtaining the maximally coherent state \(\ket{\Phi_d}=d^{-1/2}\sum_j\ket j\) from \(\ket\psi\) is \(d\,\lmin(G_\psi)\). For a balanced source it equals \(\lmin(S)/\cN(\vphi)=d\lmin(S)\,\mu(\Psi_\vphi)\) and is attained by the forward L\"owdin map followed by a diagonal phase unitary. For equal overlaps, it equals one iff \(\ket\psi\) is \(\mu\)-maximal, that is, iff \(\ket\psi=\ket{\Psi_0}\) for \(s<0\) and iff \(\ket\psi\) is a class-(A) state for \(0<s<1\).
(ii)~\emph{Coherence \(\to\) superposition.} For any \(S\), from a maximally coherent state of an orthonormal basis to a full-support target \(\ket\phi\), \(\Pst=\min\{1,1/\tau(\phi)\}\); for a balanced target \(\Pst=\min\{1,(d-1)\cN/(d-\cN)\}\). The inverse L\"owdin map achieves \(\cN(\vphi)/\lmax(S)\) for \(\ket{\Phi_\vphi}\to\ket{\Psi_\vphi}\); for equal overlaps this is optimal iff \(\ket{\Psi_\vphi}=\ket{\Psi_0}\) or, for \(s<0\) or \(d=2\), \(\ket{\Psi_\vphi}\) is a class-(A) state.
(iii)~\emph{Changing the overlap.} Between equal-overlap reference sets with overlaps \(s\) and \(s'\), and with \(\tau_0\) and \(\tau_A\) of Eq.~\eqref{eq:range},
\begin{align}
\Pst\big(\Psi_0^{(s)}\to\Psi_0^{(s')}\big)&=\min\{1,\tau_0(s)/\tau_0(s')\},\label{eq:kappa}\\
\Pst\big(\Psi_\vphi^{(s)}\to\Psi_\vphi^{(s')}\big)&=\min\{1,\tau_A(s)/\tau_A(s')\},\label{eq:eta}
\end{align}
where \(\vphi\) is any class-(A) pattern. Equation~\eqref{eq:eta} holds for \(d=2\) and, for \(d\ge3\), whenever \(s\ge0\) or \(s'\le s\); for \(s<s'\le0\) the optimal probability is \(\tau_0(s')/\tau_0(s)\) instead.
\end{theorem}

Parts (i) and (ii) are the two factors of the lower bound of Theorem~\ref{thm:inphase}(a). The guarantee~\eqref{eq:pricebound} thus describes a route through coherence: a class-(A) state is first converted into \(\ket{\Phi_d}\) with certainty and then into the target, and by Theorem~\ref{thm:price}(ii) no protocol does better in the worst case.

The optimal operation of part (i) has a simple meaning. Its successful Kraus operator \(K\) and the operator \(K^\dagger K\) are
\begin{equation}
\begin{aligned}
K&=\sqrt{\frac{\Pst}{d}}\,\sum_j\frac{\ketbra{j}{c_j^\perp}}{x_j},\\
K^\dagger K&=\frac{\Pst}{d}\sum_j\frac{\ketbra{c_j^\perp}{c_j^\perp}}{|x_j|^2}.
\end{aligned}
\label{eq:USD}
\end{equation}
The terms of this sum are the conclusive elements of an unambiguous discrimination of the reference states, which identifies \(\ket{c_j}\) with probability \(\Pst/(d\,|x_j|^2)\) and never errs~\cite{Ivanovic1987,Dieks1988,Peres1988,Chefles1998,DuanGuo1998}. The operator \(K\) reveals only that the discrimination succeeded, not which reference state was found, and therefore keeps the superposition. For a balanced source this probability is \(\lmin(S)\) for all \(j\), and \(K\) is the forward L\"owdin map \(\sqrt{\lmin(S)}\,\cT^{-1}\) followed by a diagonal phase unitary; for equal overlaps it is the optimal unambiguous discrimination of the symmetric states \(\ket{c_j}\)~\cite{CheflesBarnett1998}, and for \(d=2\) the Ivanovic--Dieks--Peres measurement. An unambiguous discrimination of nonorthogonal states fails with nonzero probability on each \(\ket{c_j}\), here with probability \(1-\lmin(S)\); on a \(\mu\)-maximal state it never fails, because the failure amplitudes of its components cancel by destructive interference.

\begin{table}[t]
\caption{Distinguished states for equal overlaps: the states with the largest minimal weight \(\mu\), the golden states, the states that yield the maximally coherent state \(\ket{\Phi_d}\) with certainty, the balanced targets for which the inverse L\"owdin map \(\cT\) is optimal, and the worst-case conversion probability of the states with the largest \(\mu\) (Proposition~\ref{prop:mu}, Theorems~\ref{thm:inphase}--\ref{thm:bridge}, and Sec.~\ref{sec:balanced}). (A) denotes the class-(A) states; for \(d=2\), \(\ket{\Psi_0}=\ket{\Psi_+}\) and class (A) is \(\ket{\Psi_-}\).}
\label{tab:roles}
\begin{ruledtabular}
\begin{tabular}{lccc}
 & \(s<0\) & \(s>0\), \(d=2\) & \(s>0\), \(d\ge3\)\\\hline
largest \(\mu\) & \(\Psi_0\) & \(\Psi_-\) & (A)\\
golden & \(\Psi_0\) & \(\Psi_-\) & none\\
\(\Phi_d\) with certainty & \(\Psi_0\) & \(\Psi_-\) & (A)\\
\(\cT\) optimal & \(\Psi_0\), (A) & \(\Psi_\pm\) & \(\Psi_0\)\\
worst case & \(1\) & \(1\) & \(1/\tau_A\)\\
\end{tabular}
\end{ruledtabular}
\end{table}

The two L\"owdin maps thus behave asymmetrically. The forward map converts every balanced source into coherence optimally, and part (i) gives the minimal weight a sharp operational meaning: for equal overlaps of either sign, the \(\mu\)-maximal states are exactly the states that yield maximal coherence with certainty. The inverse map sends every maximally coherent state of the L\"owdin frame exactly onto a balanced state~\cite{Torun2023}, but as a protocol it is optimal only for the in-phase state and, for \(s<0\) or \(d=2\), for the class-(A) states; for every other balanced target the permutation twirl of Theorem~\ref{thm:inphase}(b) succeeds with higher probability.

Table~\ref{tab:roles} collects these results for both signs of the overlap. For \(s<0\) a single state, the in-phase state, has the largest minimal weight, reaches every state with certainty, and yields maximal coherence with certainty, as every maximally coherent state does in coherence theory; the other balanced states are weaker. For \(0<s<1\) and \(d\ge3\) these roles separate: the class-(A) states have the largest minimal weight and yield maximal coherence with certainty, but no state reaches every state with certainty, and the class-(A) states reach every full-support state with probability at least \(1/\tau_A\), their exact worst case (Theorem~\ref{thm:price}). In both cases the states with the largest minimal weight are the balanced states whose components interfere most destructively, and the sign of \(s\) decides which phase pattern does so. For \(d=2\) the sign change of Sec.~\ref{sec:equal} maps the first column of the table onto the second.

Part (iii) shows that increasing \(s\) is possible with certainty for the in-phase state but not for class-(A) states, for which Theorem~\ref{thm:inphase}(a) bounds the probability by \(\tau_A(s)/\tau_A(s')<1\) for either sign of the overlaps (for \(s<0<s'\) and \(d\ge3\) the bound is in general not attained, and for \(d\ge4\) the optimum depends on the pattern \(\vphi\)), and the reverse holds for decreasing \(s\). For \(d=2\), \(\tau_A=1/\tau_0=(1+s)/(1-s)\), which equals \(\coth|\alpha|^2\) for the coherent states \(\ket{\pm\alpha}\), and Eqs.~\eqref{eq:kappa} and~\eqref{eq:eta} reproduce the diagonal entries of Table~\ref{tab:cats}. For the golden qubit state, \(\ket{\Psi_-}\) for \(s>0\) and \(\ket{\Psi_+}\) for \(s<0\), the optimal operation that reduces \(|s|\) is an optimal state separation of the reference states~\cite{CheflesBarnettSep1998}, which also fails on each \(\ket{c_j}\) but never on the golden state.

%=====================================================================
\section{Conclusions}\label{sec:conclusion}
%=====================================================================

We have derived closed forms for the optimal probability of converting one superposition of nonorthogonal states into another in three cases: when the source or the target is permutation invariant, when there are two reference states, and when the source is a Fourier state of the qutrit and the target is any full-support state. Theurer \emph{et al.} had expressed this probability as the value of a semidefinite program~\cite{Theurer2017}, which can be evaluated numerically for one source, one target, and one set of overlaps at a time. The closed forms hold for whole families of states and overlaps at once, come with optimal protocols, and settle questions that numerics can only probe, such as which conversions are one-way and which states are incomparable. They also reveal the principle at work: superposition-free operations cannot make the components of a superposition interfere more destructively with certainty while keeping all of them. The normalized squared norm \(\tau\) measures this interference. It is a strong monotone on full-support states, it bounds the optimal probability within one reference set and between different ones, and it fixes this probability for permutation-invariant sources (Theorem~\ref{thm:inphase}).

For cat states, \(\tau\) is the mean photon number in units of \(|\alpha|^2\), and every conversion among odd cat states, even cat states, and the maximally coherent qubit is a product or ratio of hyperbolic tangents, capped at one (Table~\ref{tab:cats}). The conversion between the two cat states of one amplitude is one-way: the odd cat state becomes even with certainty by a normalized photon subtraction, the even cat state becomes odd at best with probability \(\tanh^2|\alpha|^2\), and \(N\) even cat states become odd at best with probability \(\tanh^{N+1}|\alpha|^2\), which is attained for \(N\le3\) (Proposition~\ref{prop:copies}), whereas the distillation of coherence is multiplicative in the number of copies (Corollary~\ref{cor:copies}). Among the cat states with a relative phase, the odd cat state reaches all others with certainty, and the even and the Yurke--Stoler cat states are incomparable (Proposition~\ref{prop:qubit} and Corollary~\ref{cor:phase}). Because every operation that maps each of the coherent states \(\ket{\pm\alpha}\) to a coherent state is a superposition-free map (Sec.~\ref{sec:FO}), these values limit what all such operations can achieve.

Inverse L\"owdin orthogonalization~\cite{Torun2023} supplies the structure behind these results: it transports incoherent Kraus operators to the nonorthogonal reference states, maps the maximally coherent states onto the balanced states, and, through its spectrum, fixes which states have the largest minimal weight. The maximally coherent states, which are mutually equivalent and play every distinguished role in coherence theory, have no equivalent counterpart: the balanced states are inequivalent for either sign of the overlap (Fig.~\ref{fig:graphs} and Table~\ref{tab:roles}). For equal overlaps, the states with the largest minimal weight, and only they, yield a maximally coherent state with certainty. For positive overlaps and \(d\ge3\) they are nevertheless not universal sources: no pure state yields every balanced state with certainty, and their worst case, \({(d-1)(1-s)/(d-1+s)}\), is that of a route through maximal coherence. It is attained for a reference state with its in-phase part removed and meets the value one of negative overlaps with a kink at \(s=0\). Interference cannot be redistributed among the components with certainty, and the minimal weight, which decides whether maximal coherence can be extracted with certainty, is not a complete figure of merit for superposition.

Lemma~\ref{lem:branches}(a),(b), Proposition~\ref{prop:mu}, Theorems~\ref{thm:sdp} and~\ref{thm:inphase}, and the general statements of Theorem~\ref{thm:bridge}(i),(ii) hold for every Gram matrix, Lemma~\ref{lem:fourier} for every circulant one, including the coherent-state constellations of multicomponent cat codes~\cite{Grimsmo2020}, and Corollary~\ref{cor:qubit}, Proposition~\ref{prop:qubit}, and Corollary~\ref{cor:phase} for every pair of reference states once a phase of \(\ket{c_1}\) makes their overlap real. The remaining results use equal overlaps, and all results reduce to those of coherence theory for orthonormal reference states. Several questions remain open: whether the bound of Proposition~\ref{prop:copies} is attained for four copies at every overlap and for more than four copies, and the asymptotic conversion rates; the state with the largest worst-case probability when no golden state exists; the program~\eqref{eq:primal} for Gram matrices without permutation or cyclic symmetry; conversions between mixed states; and the experimental implementation of the optimal filters beyond the photon subtraction found here for cat states.

\medskip
\noindent\textbf{Use of artificial intelligence.} The author conceived the research questions and the approach, directed all stages of the work, and verified all results. The AI assistant Claude (Anthropic) was used under the author's direction to explore and check derivations and proof strategies, to write and run the numerical tests, to prepare the figures, and to improve the wording. The author is solely responsible for the content of this paper.

%\section*{Data availability}
%The data and code that support the findings of this article are available from the author upon reasonable request.

%=====================================================================
\appendix
%=====================================================================

\section{Proofs for Sec.~\ref{sec:setting}}\label{app:setting}

\begin{proof}[Proof of Eqs.~\eqref{eq:lowdinFO} and~\eqref{eq:lowdinFOcomplete}]
By Eq.~\eqref{eq:lowdin}, in the L\"owdin frame \(\ket{c_k}\) is represented by the \(k\)th column of \(S^{1/2}\) and \(\bra{c_k^\perp}\) by the \(k\)th row of \(S^{-1/2}\). The matrix \(\Pi_hD_\xi\) has the entry \(\xi_k\) in row \(h(k)\) of column \(k\) and zeros elsewhere, so the operator~\eqref{eq:FOkraus} is represented by \(S^{1/2}\Pi_hD_\xi S^{-1/2}\). Consequently \(\sum_n[K_n]_{\rm L}^\dagger[K_n]_{\rm L}=S^{-1/2}\big(\sum_nD_{\xi_n}^\dagger\Pi_{h_n}^TS\Pi_{h_n}D_{\xi_n}\big)S^{-1/2}\), which equals \(\one\) iff the bracket equals \(S\).
\end{proof}

\begin{proof}[Proof of Lemma~\ref{lem:branches}]
Parts (a) and (b) use the steps of the proof of Gram majorization in Ref.~\cite{Torun2026b}. (a) We have \(K\ket\psi=\sum_j\big(\sum_{k\in h^{-1}(j)}\xi_kx_k\big)\ket{c_j}\). Since the \(\ket{c_j}\) are linearly independent and \(K\ket\psi=z\sum_jt_j\ket{c_j}\) with all \(t_j\neq0\), every \(j\) has a nonempty preimage under \(h\). Hence \(h\) is onto, and therefore a permutation \(\pi\), and \(\xi_kx_k=zt_{\pi(k)}\). The operator \(z\,\cX_\phi U_\pi\cX_\psi^{-1}\) maps \(\ket{c_k}\) to \(z(t_{\pi(k)}/x_k)\ket{c_{\pi(k)}}=\xi_k\ket{c_{\pi(k)}}\), as \(K\) does, so the two operators agree on a basis. (b) Each \(L_\kappa\) maps \(\ket{c_j}\) to the multiple \(\braket{\kappa}{c_j}\ket{c_0}\) of a reference state and is therefore superposition-free, and \(\sum_\kappa L_\kappa^\dagger L_\kappa=\sum_\kappa\ketbra{\kappa}{\kappa}=\one-\sum_nK_n^\dagger K_n\). (c) A unitary maps linearly independent vectors to linearly independent vectors, so the index map \(h\) of a unitary FO is injective and hence a permutation. Unitarity then requires \(\bar\xi_j\xi_kS_{h(j)h(k)}=S_{jk}\) for all \(j,k\). The diagonal entries give \(|\xi_j|=1\). For \(j\neq k\) both overlaps equal \(s\neq0\), so \(\bar\xi_j\xi_k=1\), and all \(\xi_k\) equal a common phase.
\end{proof}

\section{Proof of Theorem~\ref{thm:images}}\label{app:images}

\begin{proof}[Proof of Theorem~\ref{thm:images}]
(i) By Eq.~\eqref{eq:Tequal}, \(\cT\ket{\Phi_\vphi}=q\ket{\Phi_\vphi}+(p-q)\braket{f_0}{\Phi_\vphi}\ket{f_0}=q\ket{\Phi_\vphi}+(p-q)\zeta\ket{f_0}\). Projecting onto \(\ket{l_j}\), with \(\braket{l_j}{f_0}=d^{-1/2}\), and dividing by the norm gives Eq.~\eqref{eq:Lcoeff}. The squared norm is \(q^2(1-|\zeta|^2)+p^2|\zeta|^2=\lambda_1+ds|\zeta|^2\), because \(\lambda_0-\lambda_1=ds\), and \(|\sum_je^{i\varphi_j}|^2=d+2\sum_{j<k}\cos(\varphi_j-\varphi_k)\) gives the second form.

(ii) \(\cT\ket{\Phi_\vphi}\) is proportional to \(\ket{\Phi_\vphi}\) iff \((p-q)\zeta\ket{f_0}\) is. For \(s\neq0\) we have \(p\neq q\), so either \(\zeta=0\) or \(\ket{\Phi_\vphi}\propto\ket{f_0}\), that is, all phases are equal. The eigenvalues \(q\) and \(p\) follow from Eq.~\eqref{eq:Tequal}.

(iii) By Eq.~\eqref{eq:Lcoeff}, \(\ket{\Psi_\vphi}\) is maximally coherent in the L\"owdin frame iff the numbers \(b_j=qe^{i\varphi_j}+b\), with \(b=(p-q)\zeta\), have equal moduli. Since \(|b_j|^2=q^2+|b|^2+2q\,{\rm Re}(\bar b\,e^{i\varphi_j})\), this holds iff \({\rm Re}(\bar b\,e^{i\varphi_j})\) is the same for all \(j\). If \(b=0\), this is class (A). Otherwise the points \(e^{i\varphi_j}\) lie on a line, which meets the unit circle at most twice, so the phases take at most two values. One value is class (B). For two values, after a permutation and a global phase, \(d_0\) phases equal \(0\) and \(d-d_0\) phases equal \(\varphi\neq0\), with \(1\le d_0\le d-1\); then \(b=(p-q)[d_0+(d-d_0)e^{i\varphi}]/d\), and
\begin{equation}
{\rm Re}\big[\bar b\,(1-e^{i\varphi})\big]=\frac{p-q}{d}\,(2d_0-d)(1-\cos\varphi),
\label{eq:twoblock}
\end{equation}
which vanishes iff \(d_0=d/2\), because \(p\neq q\) and \(\cos\varphi\neq1\); this requires \(d\) to be even. For \(d_0=d/2\) and \(\varphi=\pi\) we would have \(\zeta=0\), which is class (A), so \(\varphi\notin\{0,\pi\}\), and this is class (C). In this case \(\zeta=\cos\frac\varphi2\,e^{i\varphi/2}\), and the two values of \(b_j\) are \(e^{i\varphi/2}\,(p\cos\frac\varphi2\mp iq\sin\frac\varphi2)\). Thus the two blocks keep their pattern, and the relative phase \(\varphi'\) of the L\"owdin coefficients obeys \(\tan(\varphi'/2)=(q/p)\tan(\varphi/2)\), which is Eq.~\eqref{eq:contraction}.

\textit{Class (A) for \(d=3\) and \(d=4\).} Three unit phasors with zero sum form an equilateral triangle, which gives \(\ket{f_1}\) and \(\ket{f_2}\). For four unit phasors \(e_j=e^{i\varphi_j}\) with \(e_0+e_1+e_2+e_3=0\), either \(e_0+e_1=0\), and then \(e_1=-e_0\) and \(e_3=-e_2\), or \(e_0+e_1\neq0\). In the second case \(e_2\) lies on the unit circle about the origin and on the unit circle about \(-(e_0+e_1)\), because \(e_3=-(e_0+e_1)-e_2\) has unit modulus. These circles meet in at most two points, and \(-e_0\) and \(-e_1\) are such points; hence \(\{e_2,e_3\}=\{-e_0,-e_1\}\). In both cases the phasors form two antipodal pairs, which gives the circles in Table~\ref{tab:classes}.
\end{proof}

\section{Proofs for Sec.~\ref{sec:sdp}}\label{app:sdp}

\begin{proof}[Proof of Theorem~\ref{thm:sdp}]
Let \(K_\pi=\cX_\phi U_\pi\cX_\psi^{-1}\); it maps \(\ket\psi\) to \(\cX_\phi U_\pi\sum_j\ket{c_j}=\ket\phi\). By Lemma~\ref{lem:branches}(a) every successful Kraus operator has the form \(K_n=z_nK_{\pi_n}\), and it succeeds with probability \(|z_n|^2\). Grouping the operators by their permutation and setting \(w_\pi=\sum_{n:\pi_n=\pi}|z_n|^2\), the probability with which branch \(\pi\) succeeds, the success probability is \(\sum_\pi w_\pi\), and \(\sum_nK_n^\dagger K_n=\sum_\pi w_\pi K_\pi^\dagger K_\pi\). Since \(\cX_\psi\) is invertible, this operator is bounded by \(\one\) iff \(\sum_\pi w_\pi\|K_\pi\cX_\psi\ket u\|^2\le\|\cX_\psi\ket u\|^2\) for all \(\ket u=\sum_jy_j\ket{c_j}\). The right-hand side is \(y^\dagger G_\psi y\) by Eq.~\eqref{eq:Gform}, and \(K_\pi\cX_\psi\ket u=\cX_\phi U_\pi\ket u=\sum_jt_{\pi(j)}y_j\ket{c_{\pi(j)}}\) has the coefficient vector \(D_t\Pi_\pi y\), \(D_t=\diag(t)\), so that \(\|K_\pi\cX_\psi\ket u\|^2=y^\dagger\Pi_\pi^TG_\phi\Pi_\pi y\). Hence the operator inequality is equivalent to Eq.~\eqref{eq:primal}. Conversely, every feasible \(w\) defines the stochastic FO \(\{\sqrt{w_\pi}K_\pi\}\), which Lemma~\ref{lem:branches}(b) completes. The program~\eqref{eq:primal} maximizes a linear function under a linear matrix inequality, and every sufficiently small \(w>0\) is strictly feasible because \(G_\psi>0\). Strong duality therefore holds~\cite{BoydVandenberghe2004} and gives Eq.~\eqref{eq:dual}. Equation~\eqref{eq:testvector} is the inequality \(\sum_\pi w_\pi y^\dagger\Pi_\pi^TG_\phi\Pi_\pi y\le y^\dagger G_\psi y\) with every term in the sum bounded from below by its minimum.
\end{proof}

\begin{proof}[Proof of Theorem~\ref{thm:inphase}]
We use that a matrix that commutes with the permutation matrices of a doubly transitive group, in particular a permutation-invariant matrix, is a linear combination of \(\one\) and \(J\)~\cite{DixonMortimer1996}. If a component Gram matrix \(G\) has this form, its eigenvalue on \(\ones\) is \(\frac1d\ones^TG\ones=\frac1d\), and its eigenvalue on \(\ones^\perp\) is \((\tr G-\frac1d)/(d-1)\), which equals \(\tau/d\) by Eq.~\eqref{eq:tau}. Relabeling preserves the trace and the sum of all entries of a matrix. The twirl \(\langle G_\phi\rangle=|\mathcal G|^{-1}\sum_{g\in\mathcal G}\Pi_g^TG_\phi\Pi_g\) of \(G_\phi\) over a doubly transitive group \(\mathcal G\) commutes with every \(\Pi_g\) and is therefore
\begin{equation}
\langle G_\phi\rangle=\frac1d\Big[\frac Jd+\tau(\phi)\Big(\one-\frac Jd\Big)\Big].
\label{eq:twirlG}
\end{equation}

(a) \emph{Upper bound.} Let \(\{K_n\}\) be a stochastic FO whose outcomes \(\ket{\psi_n}=K_n\ket\psi/\sqrt{P_n}\) all have full support. By Lemma~\ref{lem:branches}(a), \(K_n=\sqrt{P_n}\,\cX_{\psi_n}U_{\pi_n}\cX_\psi^{-1}\) up to a phase, and the computation in the proof of Theorem~\ref{thm:sdp} turns \(\sum_nK_n^\dagger K_n\le\one\) into \(\sum_nP_n\Pi_{\pi_n}^TG_{\psi_n}\Pi_{\pi_n}\le G_\psi\). Every component Gram matrix \(G\) has \(\tr[(\one-J/d)G]=(d-1)\tau/d\), and so have its relabelings. The trace of the inequality against \(\one-J/d\ge0\) therefore gives \(\sum_nP_n\tau(\psi_n)\le\tau(\psi)\). Outcomes without full support may be present as well, because omitting them only decreases \(\sum_nK_n^\dagger K_n\), and a source without full support has no full-support outcome; hence \(\tau\), set to zero for states without full support, is a strong monotone. If \(\ket{\psi_n}=\ket\phi\) for all \(n\), the inequality reads \(\Pst\,\tau(\phi)\le\tau(\psi)\); equivalently, \(Y=d\,(\one-J/d)/[(d-1)\tau(\phi)]\) is feasible in Eq.~\eqref{eq:dual}. \emph{Lower bound.} Equal probabilities on the branches of \(\mathcal G\), with total \(W\), contribute \(W\langle G_\phi\rangle\le W\max\{1,\tau(\phi)\}\,\one/d\) to Eq.~\eqref{eq:primal}, and \(G_\psi\ge\lmin(G_\psi)\one\), so \(W=d\,\lmin(G_\psi)\min\{1,1/\tau(\phi)\}\) is feasible.

(b) By (a), \(\Pst\le\min\{1,\tau(\psi)/\tau(\phi)\}\). The permutation-invariant \(G_\psi\) has the eigenvalues \(\frac1d\) on \(\ones\) and \(\tau(\psi)/d\) on \(\ones^\perp\), so by Eq.~\eqref{eq:twirlG} the constraint \(W\langle G_\phi\rangle\le G_\psi\) reads \(W\le1\) and \(W\tau(\phi)\le\tau(\psi)\), and \(W=\min\{1,\tau(\psi)/\tau(\phi)\}\) is feasible. Only the trace and the entry sum of \(G_\phi\) entered, so the target reference set is arbitrary.

(c) For a permutation-invariant \(G_\phi\), all branches contribute the same matrix, and Eq.~\eqref{eq:primal} reduces to \((\sum_\pi w_\pi)G_\phi\le G_\psi\). Hence \(\Pst=\lmin(G_\psi,G_\phi)\), attained by the identity branch alone.

\textit{Invariant states.} A permutation-invariant \(G_\psi\) has equal diagonal entries \(|x_j|^2\) and equal off-diagonal entries \(\bar x_jS_{jk}x_k=(G_\psi)_{01}\), which are real because \(G_\psi\) is Hermitian. With \(x_j=|x_j|e^{i\varphi_j}\), this gives \(S_{jk}=[(G_\psi)_{01}/|x_0|^2]\,e^{i(\varphi_j-\varphi_k)}\) for \(j\neq k\), the rephasing statement after Theorem~\ref{thm:inphase}. For equal overlaps \(s\neq0\) and balanced \(x_j\propto e^{i\varphi_j}\), the phases \(e^{i(\varphi_k-\varphi_j)}\), \(j\neq k\), are then all equal to the same real number, \(+1\) or \(-1\); for \(d\ge3\), \(-1\) is impossible, because \(\varphi_1-\varphi_0=\varphi_2-\varphi_0=\pi\) would give \(\varphi_2-\varphi_1=0\), so all phases are equal.
\end{proof}

\begin{proof}[Proof of Eqs.~\eqref{eq:range}, \eqref{eq:from}, and~\eqref{eq:fromBSS}]
For \(x\propto\ones\), \(\|x\|^2=1/\lambda_0\) and \(\tau=(d/\lambda_0-1)/(d-1)=\lambda_1/\lambda_0\), because \(d-\lambda_0=(d-1)\lambda_1\); for \(x\perp\ones\), \(\|x\|^2=1/\lambda_1\) and \(\tau=(d/\lambda_1-1)/(d-1)=\tau_A\). These are the extreme values of \(\|x\|^2\), because \(x^\dagger Sx=1\), and \(\lmax(S)=\lambda_0\) for \(s>0\) and \(\lambda_1\) for \(s<0\). Part (b) with \(\tau(\Psi_0)=\tau_0\) is Eq.~\eqref{eq:from}. For \(s\le0\), \(\tau(\phi)\le\tau_0\) and \(\Pst=1\); for \(s>0\), \(\tau(\phi)\ge\tau_0\) with equality iff \(t\propto\ones\). For balanced targets \(\|t\|^2=1/\cN\) and \(\tau=(d-\cN)/[(d-1)\cN]\), which gives Eq.~\eqref{eq:fromBSS}.
\end{proof}

\begin{proof}[Proof of Eq.~\eqref{eq:to}]
For \(\phi=\Psi_0\), \(G_\phi=S/(d\lambda_0)\), and part (c) gives \(\Pst=d\lambda_0\lmin(G_\psi,S)\), attained by the filter \(\sqrt{\Pst}\,\cX_{\Psi_0}\cX_\psi^{-1}=\sqrt{\Pst/(d\lambda_0)}\,\cX_\psi^{-1}\). By Eq.~\eqref{eq:Gform}, \(\lmin(G_\psi,S)=\min_u\|\cX_\psi\ket u\|^2/\|\ket u\|^2=\lmin(\cX_\psi^\dagger\cX_\psi)\). For a class-(A) source, \(x=e_\vphi/\sqrt{d\lambda_1}\), and \(G_\psi\) is given by Eq.~\eqref{eq:GclassA}. On the orthonormal pair \(\ones/\sqrt d\), \(v/\sqrt d\) the pencil \((d\lambda_1G_\psi,S)\) is \(\diag(\lambda_1,\lambda_0)\) versus \(\diag(\lambda_0,\lambda_1)\), and on the orthogonal complement both matrices equal \(\lambda_1\one\). Therefore \(\Pst=(\lambda_0/\lambda_1)\min\{1,\lambda_1/\lambda_0,\lambda_0/\lambda_1\}\), which is \(1\) for \(s>0\) and \((\lambda_0/\lambda_1)^2\) for \(s<0\).
\end{proof}

\begin{proof}[Proof of Eq.~\eqref{eq:secondorder}]
The left-hand side is \(\lmin(G_\psi,S)\) divided by \(|x_{j^\ast}|^2\), which is invariant under rescaling of \(x\), so the normalization may be dropped. Write \(S=\one+sE\) with \(E=J-\one\), so that \(G_\psi=X^\dagger X+s\,X^\dagger EX\). At \(s=0\) the pencil \((X^\dagger X,\one)\) has the simple smallest eigenvalue \(\mu_\ast=|x_{j^\ast}|^2\), whose eigenvector is the \(j^\ast\)th coordinate vector. Its first-order shift is \((X^\dagger EX-\mu_\ast E)_{j^\ast j^\ast}=0\), because \(E\) has zero diagonal. Second-order perturbation theory gives the shift \(-s^2\sum_{k\neq j^\ast}|(X^\dagger EX-\mu_\ast E)_{kj^\ast}|^2/(|x_k|^2-\mu_\ast)\), in which \((X^\dagger EX-\mu_\ast E)_{kj^\ast}=x_{j^\ast}(\bar x_k-\bar x_{j^\ast})\). Dividing by \(\mu_\ast\) gives Eq.~\eqref{eq:secondorder}.
\end{proof}

\begin{proof}[Proof of Corollary~\ref{cor:qubit} and Eq.~\eqref{eq:qubitK}]
For \(d=2\) and real \(s\), \(\ket{\Psi_+}=\ket{\Psi_0}\) has a permutation-invariant component Gram matrix, and \(\det(G_\psi-\lambda S)=(1-s^2)\lambda^2-Q_+\lambda+(1-s^2)|x_0x_1|^2\). Its smaller root, multiplied by \(d\lambda_0=2(1+s)\), is Eq.~\eqref{eq:to} and gives the upper sign of Eq.~\eqref{eq:qubit}. The sign change \(\ket{c_1}\to-\ket{c_1}\) maps \(s\to-s\), \(x_1\to-x_1\), and \(\ket{\Psi_-}\to\ket{\Psi_+}\), which gives the lower sign. The special values are \(\Pst(\Psi_\mp\to\Psi_\pm)=\min\{1,\tau(\Psi_\mp)/\tau(\Psi_\pm)\}\) by Theorem~\ref{thm:inphase}(b), with \(\tau(\Psi_\pm)=(1\mp s)/(1\pm s)\). The same formula gives Table~\ref{tab:cats}, since \((1-s)/(1+s)=\tanh|\alpha|^2\) for \(s=e^{-2|\alpha|^2}\) and \(\tau(\Phi_2)=1\). The operator~\eqref{eq:qubitK} is \(\sqrt{\lambda_1/\lambda_0}\) times the coefficient operator with \(x=(1,-1)\); in the L\"owdin frame it maps \(\ket{-_{\rm L}}\mapsto\ket{+_{\rm L}}\) and \(\ket{+_{\rm L}}\mapsto(\lambda_1/\lambda_0)\ket{-_{\rm L}}\), so its singular values are \(1\) and \(\lambda_1/\lambda_0\), and the completion follows from Lemma~\ref{lem:branches}(b). For \(s<0\) the sign change turns it into the operator \(\sqrt{\lambda_0/\lambda_1}\,(\ketbra{c_0}{c_0^\perp}-\ketbra{c_1}{c_1^\perp})\) of Sec.~\ref{sec:cats}.
\end{proof}

\begin{proof}[Proof of Proposition~\ref{prop:copies}]
The matrices \(G_{\Psi_\pm}\) have the eigenvalue \(\frac12\) on \((1,1)\) and the eigenvalue \((1\mp s)/[2(1\pm s)]=r^{\pm1}/2\) on \((1,-1)\). With the orthonormal vectors \(\chi_k=2^{-N/2}\big((-1)^{k\cdot j}\big)_j\), where \(j,k\in\mathbb Z_2^N\) and \(|k|\) denotes the number of ones in \(k\), the source and the target therefore have
\begin{equation}
G_{\Psi_\pm^{\otimes N}}=2^{-N}\sum_kr^{\pm|k|}\,\chi_k\chi_k^T .
\label{eq:walsh}
\end{equation}
\emph{Upper bound.} Take the test vector \(y=\chi_{\ones}\) in Eq.~\eqref{eq:testvector}, that is, \(k=\ones=(1,\dots,1)\). The numerator is \(2^{-N}r^N\). For every permutation, \(\Pi_\pi\chi_{\ones}\) is a unit vector with vanishing entry sum, hence orthogonal to \(\chi_0\), and on \(\chi_0^\perp\) the target matrix is at least \(2^{-N}r^{-1}\), because \(r<1\). The denominator is therefore at least \(2^{-N}r^{-1}\), and \(\Pst\le r^{N+1}\). \emph{Lower bounds.} Converting each copy with the optimal operation of Corollary~\ref{cor:qubit} succeeds with probability \(r^{2N}\). For an invertible binary matrix \(A\), let \(\pi(j)=A^Tj\) modulo 2. Since \(k\cdot A^Tj=(Ak)\cdot j\), we have \(\Pi_\pi^T\chi_k=\chi_{Ak}\), so \(\Pi_\pi^TG_\phi\Pi_\pi=2^{-N}\sum_kr^{-|k|}\chi_{Ak}\chi_{Ak}^T\) is diagonal in the same basis as \(G_\psi\), and the branch \(\pi\) alone is feasible in Eq.~\eqref{eq:primal} with probability \(\min_{k\neq0}r^{|k|+|Ak|}\). The matrices
\begin{equation}
A=\begin{pmatrix}1&1\\0&1\end{pmatrix},\quad A=\begin{pmatrix}1&1&1\\1&0&1\\0&1&1\end{pmatrix}
\label{eq:Amatrices}
\end{equation}
have \(|k|+|Ak|\le N+1\) for all \(k\neq0\), as one checks on the three and the seven nonzero \(k\), which gives \(r^{N+1}\) for \(N=2\) and \(N=3\); for \(N=1\) the bound is Corollary~\ref{cor:qubit}.
\end{proof}

\begin{proof}[Proof of Corollary~\ref{cor:copies}]
The \(d^N\) product reference states have the Gram matrix \(S^{\otimes N}\), and \(G_{\psi^{\otimes N}}=G_\psi^{\otimes N}\) has the smallest eigenvalue \(\lmin(G_\psi)^N\) and the trace \(\|x\|^{2N}\). Since \(\ket{\Phi_d}^{\otimes N}\) is the maximally coherent state of the product basis, with component Gram matrix \(\one/d^N\), Theorem~\ref{thm:inphase}(c) gives the first formula and Theorem~\ref{thm:inphase}(b) the second, with \(\tau(\phi^{\otimes N})=(d^N\|t\|^{2N}-1)/(d^N-1)\). Since \(d\|t\|^2=1+(d-1)\tau(\phi)\), two states \(\ket\phi\) and \(\ket{\phi'}\) of \(d\) and \(d'\) reference states have \((dd'-1)\,\tau(\phi\otimes\phi')=(d-1)\tau(\phi)+(d'-1)\tau(\phi')+(d-1)(d'-1)\tau(\phi)\tau(\phi')\), which is smaller than \((dd'-1)\,\tau(\phi)\tau(\phi')\) if \(\tau(\phi)>1\) and \(\tau(\phi')>1\). Hence \(\tau(\phi^{\otimes N})<\tau(\phi)^N\) for \(\tau(\phi)>1\) and \(N\ge2\), and the second probability exceeds the \(N\)th power of its value for \(N=1\).
\end{proof}

\begin{proof}[Proof of Proposition~\ref{prop:qubit} and Corollary~\ref{cor:phase}]
Let \(w_{\rm id}=(W+V)/2\) and \(w_{\rm ex}=(W-V)/2\) be the probabilities of the two branches. The constraint of Eq.~\eqref{eq:primal} is \(H\ge0\) with \(H=G_\psi-w_{\rm id}G_\phi-w_{\rm ex}G_{\rm ex}\). The set of \((W,V)\) with \(H\ge0\) is convex, and its boundary belongs to the curve \(\det H=0\). With \(\det G_{\rm ex}=\det G_\phi\), the expansion of the determinant gives
\begin{equation}
\begin{aligned}
\det H={}&\det G_\psi-\tfrac12(\Omega_{\rm id}+\Omega_{\rm ex})W-\tfrac12(\Omega_{\rm id}-\Omega_{\rm ex})V\\
&+\tfrac14\Omega_+W^2-\tfrac14\Omega_-V^2 .
\end{aligned}
\label{eq:detH}
\end{equation}
A direct computation gives \(\Omega(G_\phi,G_{\rm ex})=|t_0|^4+|t_1|^4-2s^2{\rm Re}[(\bar t_0t_1)^2]\) and \(\det G_\phi=(1-s^2)|t_0t_1|^2\), hence \(\Omega_+>0\) and the expression for \(\Omega_-\ge0\) given after the proposition. For a single branch, \(V=\pm W\), the determinant is \(\det G_\psi-\Omega_{\rm id,ex}W+\det G_\phi W^2\), and \(H\ge0\) up to its smaller root, which is \(P_{\rm id,ex}\). Let \(\Omega_->0\). For fixed \(W\) the determinant is largest at \(V_0=(\Omega_{\rm ex}-\Omega_{\rm id})/\Omega_-\), where it equals \(\frac14\Omega_+W^2-\frac12(\Omega_{\rm id}+\Omega_{\rm ex})W+\det G_\psi+(\Omega_{\rm id}-\Omega_{\rm ex})^2/(4\Omega_-)\). The largest \(W\) with \(H\ge0\) for some real \(V\) is the smaller root of this quadratic, which is \(P_{\rm mix}\), and it is attained only at \(V=V_0\). If \(|V_0|\le P_{\rm mix}\), both branch probabilities are nonnegative and \(\Pst=P_{\rm mix}\). Otherwise the maximum of \(W\) over the convex set \(\{H\ge0,\ w_{\rm id}\ge0,\ w_{\rm ex}\ge0\}\) is not attained in the interior of the quadrant, so it is attained on a single branch and is smaller than \(P_{\rm mix}\). If \(\Omega_-=0\), then \(G_{\rm ex}=G_\phi\), and both branches give Corollary~\ref{cor:qubit}.

For the balanced states, \(x=(1,e^{i\varphi})/\sqrt{2\cN(\varphi)}\) and \(t=(1,e^{i\vartheta})/\sqrt{2\cN(\vartheta)}\), one finds \(\det G_\psi=(1-s^2)/[4\cN(\varphi)^2]\), \(\det G_\phi=(1-s^2)/[4\cN(\vartheta)^2]\), \(\Omega_{\rm id,ex}=[1-s^2\cos(\varphi\mp\vartheta)]/[2\cN(\varphi)\cN(\vartheta)]\), \(\Omega_+=(1-s^2\cos^2\vartheta)/\cN(\vartheta)^2\), and \(\Omega_-=s^2\sin^2\vartheta/\cN(\vartheta)^2\). Then \(\sqrt{\Omega_0}=|s(\cos\varphi-\cos\vartheta)|/[\cN(\varphi)\cN(\vartheta)]\), and
\begin{equation}
P_{\rm mix}=\frac{1-s^2\cos\varphi\cos\vartheta-|s(\cos\varphi-\cos\vartheta)|}{(1+s\cos\varphi)(1-s\cos\vartheta)},
\label{eq:Pmixphase}
\end{equation}
which equals \(\tau(\Psi_\varphi)/\tau(\Psi_\vartheta)\) if \(s\cos\varphi\ge s\cos\vartheta\) and \(1\) otherwise, that is, \(P_{\rm mix}=P_\tau\). Moreover \(V_0=\sin\varphi\,\cN(\vartheta)/[\cN(\varphi)\sin\vartheta]\), so the condition \(|V_0|\le P_{\rm mix}\) is the condition of the corollary; for \(\sin\vartheta=0\) it holds iff \(\sin\varphi=0\), in agreement with Theorem~\ref{thm:inphase}(b). Since \(\cos(\varphi-\vartheta)\ge\cos(\varphi+\vartheta)\) for \(0\le\varphi,\vartheta\le\pi\), the better single branch is the identity, and \(P_{\rm id}\) is the second expression of the corollary. For \(\varphi=\pi/2\) and \(\vartheta=0\) it is \([1-s\sqrt{2-s^2}]/(1-s)\).
\end{proof}

\section{Proofs of Theorems~\ref{thm:nogo} and~\ref{thm:price} and Table~\ref{tab:qutrit}}\label{app:nogo}

\begin{proof}[Proof of Theorem~\ref{thm:nogo}(i)]
The source has the component Gram matrix~\eqref{eq:GclassA}, with \(v\perp\ones\) and \(\|v\|^2=d\). Let \(\Xi=\one-J/d-vv^\dagger/d\) be the projector orthogonal to \(\ones\) and \(v\), so that \(\tr(\Xi G_\psi)=(d-2)/d\). The permuted component Gram matrices of the target, \(G_\pi=\Pi_\pi^TG_\phi\Pi_\pi\), have \(\tr G_\pi=\|t\|^2\), \(\ones^TG_\pi\ones=1\), and, by Eq.~\eqref{eq:Gform},
\begin{equation}
v^\dagger G_\pi v=\Big\|\sum_jt_{\pi(j)}e^{-i\varphi_j}\ket{c_{\pi(j)}}\Big\|^2=\lambda_1\|t\|^2+s\,a_\pi .
\label{eq:vAv}
\end{equation}
Their mean over all permutations is the twirl~\eqref{eq:twirlG} with \(\mathcal G=\Sym_d\), for which \(v^\dagger\langle G_\phi\rangle v=\tau(\phi)\) and \(d\,\tr(\Xi\langle G_\phi\rangle)=(d-2)\,\tau(\phi)\). Hence \(\lambda_1\|t\|^2+s\,\langle a\rangle=\tau(\phi)\), which is the expression for \(\langle a\rangle\) given before the theorem, and
\begin{equation}
d\,\tr(\Xi G_\pi)=(d-2)\,\tau(\phi)-s\,(a_\pi-\langle a\rangle).
\label{eq:trQ}
\end{equation}
The trace of Eq.~\eqref{eq:primal} against \(\Xi\ge0\) gives \(\sum_\pi w_\pi\,d\,\tr(\Xi G_\pi)\le d-2\), which implies Eq.~\eqref{eq:witness}; the right-hand side of Eq.~\eqref{eq:trQ} is positive because \(G_\pi>0\) and \(\Xi\neq0\). For a balanced target, \(\|t\|^2=1/\cN(\vth)\) and \(a_\pi=d\,|\zeta_\pi|^2/\cN(\vth)\), so that \(d\,\tr(\Xi G_\pi)=(d-1+s)\|t\|^2-1-s\,a_\pi\) gives Eq.~\eqref{eq:witnessBSS}. For a class-(A) target, \(\cN(\vth)=\lambda_1\), and this bound is below one iff \(\max_\pi|\zeta_\pi|<1\), that is, iff \(\vth\) is not a permutation of \(\vphi\) up to a global phase. For \(d=3\), \(\vphi\) the pattern of \(\ket{f_1}\), and \(\vth=(0,\pi,\pi)\), every permutation gives \(|\zeta_\pi|^2=4/9\), and with \(\cN(\vth)=1-2s/3\) the bound is \((3-2s)/(3+s)\).
\end{proof}

\begin{lemma}[Fourier sources]
\label{lem:fourier}
Let \(S\) be circulant, \(S_{jk}=S_{0,k-j}\), with eigenvalues \(\Lambda_k\) on the vectors \((\omega^{jk})_j\), and let \(\ket{f_k}\propto\sum_j\omega^{jk}\ket{c_j}\); these states form an orthonormal basis and coincide with the Fourier states of the L\"owdin frame. For the source \(\ket{f_m}\) and any full-support target,
\begin{equation}
\Pst(f_m\to\phi)=\max_{w\ge0}\Big\{\textstyle\sum_\pi w_\pi:\ \sum_\pi w_\pi\,\nu_k(\pi)\le1\ \forall k\Big\},
\label{eq:LP}
\end{equation}
where \(\nu_k(\pi)=\|\cX_\phi U_\pi\cX_{f_m}^{-1}\ket{f_k}\|^2\) is the ratio of the probability with which branch \(\pi\) is triggered by \(\ket{f_k}\) to the probability with which it succeeds on \(\ket{f_m}\),
\begin{equation}
\nu_k(\pi)=\frac{\Lambda_m}{\Lambda_k}\Big\|\sum_j\omega^{j(k-m)}t_{\pi(j)}\ket{c_{\pi(j)}}\Big\|^2 .
\label{eq:LPcoef}
\end{equation}
The coefficients \(\nu_k(\pi)\) depend on \(\pi\) only through its coset modulo cyclic shifts.
\end{lemma}

\begin{proof}
For circulant \(S\) the cyclic shift \(\sigma:j\mapsto j+1\) preserves \(S\), so \(U_\sigma\) is a free unitary; its eigenvectors \(\ket{f_k}\) are orthonormal, with distinct eigenvalues. Since \(\cX_{f_m}^{-1}=\sqrt{d\Lambda_m}\sum_j\omega^{-jm}\ketbra{c_j}{c_j^\perp}\), we have \(U_\sigma\cX_{f_m}^{-1}=\omega^m\cX_{f_m}^{-1}U_\sigma\), and therefore \(K_{\pi\sigma}=\omega^mK_\pi U_\sigma\) for \(K_\pi=\cX_\phi U_\pi\cX_{f_m}^{-1}\). Hence \(K_{\pi\sigma}^\dagger K_{\pi\sigma}=U_\sigma^\dagger K_\pi^\dagger K_\pi U_\sigma\), and \(\nu_k(\pi\sigma)=\nu_k(\pi)\), because \(U_\sigma\ket{f_k}\) equals \(\ket{f_k}\) up to a phase. For \(w\ge0\) let \(M=\sum_\pi w_\pi K_\pi^\dagger K_\pi\), and let \(\langle w\rangle\) be the average of \(w\) over the cosets \(\pi C_d\) of the cyclic group \(C_d\) generated by \(\sigma\), \(\langle w\rangle_\pi=\frac1d\sum_\ell w_{\pi\sigma^\ell}\). Then \(\langle w\rangle\) has the same total as \(w\), and \(\langle M\rangle=\sum_\pi\langle w\rangle_\pi K_\pi^\dagger K_\pi\) is the average of \(M\) over conjugation by the powers of \(U_\sigma\). It commutes with \(U_\sigma\) and is therefore diagonal in \(\{\ket{f_k}\}\), with the same diagonal entries \(\sum_\pi w_\pi \nu_k(\pi)\) as \(M\). If \(w\) is feasible in Eq.~\eqref{eq:primal}, then \(M\le\one\) and these entries are at most one; conversely, if they are at most one, then \(\langle M\rangle\le\one\) and \(\langle w\rangle\) is feasible with the same total. This proves Eq.~\eqref{eq:LP}, and it shows that the optimum is attained with branch probabilities that are constant on the cosets. Finally, \(\ket{f_k}=\sum_j\omega^{jk}\ket{c_j}/\sqrt{d\Lambda_k}\), so \(\cX_{f_m}^{-1}\ket{f_k}=\sqrt{\Lambda_m/\Lambda_k}\sum_j\omega^{j(k-m)}\ket{c_j}\), and applying \(\cX_\phi U_\pi\) gives Eq.~\eqref{eq:LPcoef}.
\end{proof}

\begin{proof}[Proof of Theorem~\ref{thm:nogo}(ii) and Eq.~\eqref{eq:K1R}]
The transposition of the indices \(1\) and \(2\) is a free unitary that exchanges \(\ket{f_1}\) and \(\ket{f_2}\), so \(\Pst(f_2\to\phi)=\Pst(f_1\to\phi)\), and we take the source \(\ket{f_1}\). For equal overlaps, \(\Lambda_0=\lambda_0\), \(\Lambda_1=\Lambda_2=\lambda_1\), and \(\|\sum_jy_j\ket{c_j}\|^2=\lambda_1\|y\|^2+s|\sum_jy_j|^2\) for every vector \(y\). Let \(a_\pm=|\sum_j\omega^{\mp j}t_j|^2\). The alignment~\eqref{eq:align} is \(a_+\) for the three even (cyclic) permutations and \(a_-\) for the three odd ones, so that \(\langle a\rangle=(a_++a_-)/2\) and \(a_{\max}-\langle a\rangle=|a_+-a_-|/2\), which we denote by \(\Gamma\). Let \(a_+\ge a_-\); otherwise exchange the even and the odd branches below. Since \(\lambda_1\|t\|^2+s\,\langle a\rangle=\tau(\phi)\), Eq.~\eqref{eq:LPcoef} with \(m=1\) gives \(\nu_1=1\) for every \(\pi\), \(\nu_2=\tau(\phi)-s\Gamma\) and \(\nu_0=(\lambda_1/\lambda_0)[\tau(\phi)+s\Gamma]\) for even \(\pi\), and the same expressions with \(\Gamma\to-\Gamma\) for odd \(\pi\). With \(w_{\rm e}\) and \(w_{\rm o}\) the total probabilities of the even and of the odd branches and \(\tau=\tau(\phi)\), Eq.~\eqref{eq:LP} is the maximum of \(w_{\rm e}+w_{\rm o}\) under
\begin{equation}
\begin{aligned}
w_{\rm e}+w_{\rm o}&\le1,\\
(\tau-s\Gamma)\,w_{\rm e}+(\tau+s\Gamma)\,w_{\rm o}&\le1,\\
(\tau+s\Gamma)\,w_{\rm e}+(\tau-s\Gamma)\,w_{\rm o}&\le\lambda_0/\lambda_1.
\end{aligned}
\label{eq:qutritLP}
\end{equation}
The first two constraints give \(w_{\rm e}+w_{\rm o}\le W=\min\{1,1/(\tau-s\Gamma)\}\). Conversely, for \(w_{\rm e}+w_{\rm o}=W\) the last two constraints read \((\tau+s\Gamma)W-\lambda_0/\lambda_1\le2s\Gamma\,w_{\rm o}\le1-(\tau-s\Gamma)W\), and some \(w_{\rm o}\in[0,W]\) satisfies them, because \(2\tau W\le2\tau_A=1+\lambda_0/\lambda_1\) by Eq.~\eqref{eq:range} and \((\tau-s\Gamma)W\le1<\lambda_0/\lambda_1\). Hence \(\Pst=\min\{1,1/(\tau-s\Gamma)\}\), the bound~\eqref{eq:witness} capped at one. To express it by populations, note that \(\braket{f_k}{\phi}=\sqrt{\Lambda_k/3}\,\sum_j\omega^{-jk}t_j\), so that \(a_+=3\rho_1/\lambda_1\), \(a_-=3\rho_2/\lambda_1\), and \(|\sum_jt_j|^2=3\rho_0/\lambda_0\). Then \(\|t\|^2=\rho_0/\lambda_0+(\rho_1+\rho_2)/\lambda_1\), which with \(\rho_0+\rho_1+\rho_2=1\) gives \(\tau=1+3s\,[(\rho_1+\rho_2)/(2\lambda_1)-\rho_0/\lambda_0]\), and \(s\Gamma=3s\,|\rho_1-\rho_2|/(2\lambda_1)\); their difference is the bracket of Eq.~\eqref{eq:qutritall}. For \(\Gamma=0\), \(w_{\rm o}=0\) is a solution: equal probabilities on the three cyclic branches \(\pi=\sigma^\ell\), \(\ell=0,1,2\). The target \(\ket R\) has the real coefficients \(t=\varepsilon/\sqrt{3\cN_R}\) with \(3\cN_R=3-2s\), so \(\Gamma=0\) and \(\tau(R)=(3+s)/(3-2s)>1\); the operators \(\sqrt{\Pst/3}\,\cX_RU_{\sigma^\ell}\cX_{f_1}^{-1}\) are Eq.~\eqref{eq:K1R}, and the binding constraint, \(k=2\), is the test-vector bound~\eqref{eq:testvector} with the test state \(\ket{f_2}\).
\end{proof}

\begin{proof}[Proof of Theorem~\ref{thm:price}]
(i)~By Eq.~\eqref{eq:GclassA}, \(\lmin(G_\psi)=1/d\) for \(s>0\), so the lower bound of Theorem~\ref{thm:inphase}(a) is \(\min\{1,1/\tau(\phi)\}\). It equals \(\Pst(\Phi_d\to\phi)\) by Theorem~\ref{thm:inphase}(b), because \(G_{\Phi_d}=\one/d\) is permutation invariant with \(\tau=1\), and \(\tau(\phi)\le\tau_A\) by Eq.~\eqref{eq:range}.

(ii)~If \(a_\pi=\langle a\rangle\) for all \(\pi\), Eq.~\eqref{eq:witness} gives \(\Pst\le1/\tau(\phi)\), and \(\Pst\le1\) always. Let all coefficients but \(t_k\) be equal to \(t_{k'}\), \(k'\neq k\). Since \(\sum_je^{-i\varphi_j}=0\), \(\sum_jt_{\pi(j)}e^{-i\varphi_j}=(t_k-t_{k'})\,e^{-i\varphi_{\pi^{-1}(k)}}\), whose modulus does not depend on \(\pi\). The state~\eqref{eq:hardest} has \(t_k\propto d-1\) and \(t_j\propto-1\) for \(j\neq k\), so \(t\) has full support and is orthogonal to \(\ones\), and \(\tau(\phi_k)=\tau_A\). Moreover \(\ket{\phi_k}\propto\ket{c_k}-\frac1d\sum_j\ket{c_j}=(\one-\ketbra{\Psi_0}{\Psi_0})\ket{c_k}\), because \(\braket{\Psi_0}{c_k}=\sqrt{\lambda_0/d}\).

\textit{Two-component targets.} Let \(\sum_je^{2i\varphi_j}=0\), that is, \(v^Tv=0\), so that \(\bar v\) is orthogonal to \(v\) and to \(\ones\). With the projector \(\Xi\) of the proof of Theorem~\ref{thm:nogo}(i), let
\begin{equation}
Y=\frac{d}{(d-1)\,\tau_A}\Big(\Xi+\frac{\bar v\,v^T}{d}\Big)\ge0.
\label{eq:Yprice}
\end{equation}
Since \(Yv=0\), \(\tr(YG_\psi)=\tr Y/d=1/\tau_A\). Let \(t\) be the real coefficient vector of \((\ket{c_j}-\ket{c_k})/\sqrt{2\lambda_1}\), \(j\neq k\), with the entries \(t_j=-t_k=1/\sqrt{2\lambda_1}\) and zeros elsewhere. Then \(G_\phi=\lambda_1\diag(t)^2+s\,tt^T\), and every \(\Pi_\pi^TG_\phi\Pi_\pi\) has the same form for another pair of indices. The diagonal entries of \(Y\) all equal \(1/\tau_A\), and \(t^TYt=d/[(d-1)\tau_A\lambda_1]\), because \(|v^Tt|=|v^\dagger t|\) for real \(t\). Hence \(\tr(Y\Pi_\pi^TG_\phi\Pi_\pi)=[1+sd/((d-1)\lambda_1)]/\tau_A=1\) for every \(\pi\). For a sequence of full-support targets whose coefficient vectors tend to \(t\), the smallest of these traces tends to one by continuity, so Eq.~\eqref{eq:dual} bounds the limit superior of \(\Pst\) by \(1/\tau_A\), and part (i) is the matching lower bound. For \(\ket{f_1}\), \(v=(\omega^{-j})_j\) and \(v^Tv=\sum_j\omega^{-2j}=0\) for \(d\ge3\).
\end{proof}

\begin{proof}[Proof of Table~\ref{tab:qutrit} and Eqs.~\eqref{eq:K01} and~\eqref{eq:KR1}]
The \(\Psi_0\) entries follow from Eq.~\eqref{eq:fromBSS} with \(\cN(f_1)=\lambda_1\) and \(\cN(R)=\cN_R=1-2s/3\); Eq.~\eqref{eq:K01} is the twirl~\eqref{eq:twirl} with \(\cX_{\Psi_0}=\one/\sqrt{3\lambda_0}\), \(\cX_{f_1}=Z_\omega/\sqrt{3\lambda_1}\), and the doubly transitive group \(\mathcal G=\Sym_3\). The entry \(f_1\to R\) is Theorem~\ref{thm:nogo}(ii). For \(R\to\Psi_0\), the matrices \(3\cN_RG_R=(1-s)\one+s\,\varepsilon\varepsilon^T\) and \(S=(1-s)\one+sJ\) both equal \((1-s)\one\) on the orthogonal complement of \({\rm span}\{\ones,\varepsilon\}\), where the pencil \((3\cN_RG_R,S)\) has the eigenvalue \(1\). In the orthonormal basis \(\{\ones/\sqrt3,(3\varepsilon+\ones)/(2\sqrt6)\}\) of \({\rm span}\{\ones,\varepsilon\}\), in which \(\varepsilon=(-1/\sqrt3,\sqrt{8/3})\), the two matrices are
\begin{equation}
\begin{pmatrix}1-\frac{2s}{3}&-\frac{2\sqrt2\,s}{3}\\-\frac{2\sqrt2\,s}{3}&1+\frac{5s}{3}\end{pmatrix},\quad\begin{pmatrix}\lambda_0&0\\0&\lambda_1\end{pmatrix}.
\label{eq:pencilR}
\end{equation}
Their generalized eigenvalues have the product \(1\) and the sum \(2(1+s+2s^2)/(\lambda_0\lambda_1)\), so they are \(\lambda_\pm=(\sqrt{1+s}\pm\sqrt2\,s)^2/(\lambda_0\lambda_1)\), with \(\lambda_-\le1\). Hence \(\lmin(G_R,S)=\lambda_-/(3\cN_R)\), and Eq.~\eqref{eq:to} gives \(\Pst(R\to\Psi_0)=\lambda_0\lambda_-/\cN_R\), the entry of Table~\ref{tab:qutrit}. For \(R\to f_1\), Eq.~\eqref{eq:GclassA} gives \(G_{f_1}=[(1-s)\one+s\,vv^\dagger]/(3\lambda_1)\) with \(v=(\omega^{-j})_j\), and \(\Pi_\pi^Tvv^\dagger\Pi_\pi\) equals \(vv^\dagger\) for even and \(\bar v\bar v^\dagger\) for odd \(\pi\). Let \(w_{\rm e}\) and \(w_{\rm o}\) be the total probabilities of the even and odd branches, and \(W=w_{\rm e}+w_{\rm o}\). In the orthonormal basis \(\{\ones,v,\bar v\}/\sqrt3\), in which \(\varepsilon=(-1,2,2)/\sqrt3\), the constraint~\eqref{eq:primal}, multiplied by \(3\), becomes
\begin{equation}
\begin{aligned}
&\frac{\lambda_1}{\cN_R}\one+\frac{s}{3\cN_R}\begin{pmatrix}1&-2&-2\\-2&4&4\\-2&4&4\end{pmatrix}\\
&\quad-\diag\!\Big(W,W+\tfrac{3sw_{\rm e}}{\lambda_1},W+\tfrac{3sw_{\rm o}}{\lambda_1}\Big)\ge0.
\end{aligned}
\label{eq:Rf1matrix}
\end{equation}
Positivity of the lower \(2\times2\) block requires, by the arithmetic-geometric mean inequality, \(\lambda_1/\cN_R\ge W(2+s)/(2\lambda_1)\), with equality possible only for \(w_{\rm e}=w_{\rm o}\). For \(w_{\rm e}=w_{\rm o}\) and this value of \(W\) the whole matrix is positive semidefinite, because \(W\le\lambda_1/\cN_R\). Hence \(\Pst(R\to f_1)=2\lambda_1^2/[(2+s)\cN_R]\), attained with equal probabilities on the six branches of \(\Sym_3\), which is Eq.~\eqref{eq:KR1}.
\end{proof}

\section{Proof of Theorem~\ref{thm:bridge}}\label{app:bridge}

\begin{proof}[Proof of Theorem~\ref{thm:bridge}]
(i) The target \(\ket{\Phi_d}\) has \(G_\phi=\one/d\), which is permutation invariant, and Theorem~\ref{thm:inphase}(c) gives \(\Pst=d\lmin(G_\psi)\). For a balanced source, \(G_\psi=D_\vphi^\dagger SD_\vphi/(d\cN)\) with the unitary \(D_\vphi=\diag(e_\vphi)\), so \(\lmin(G_\psi)=\lmin(S)/(d\cN)\). The operator \(\sqrt{\lmin(S)}\,\cT^{-1}\) is superposition-free from \(\{\ket{c_j}\}\) to \(\{\ket{l_j}\}\), satisfies \(\lmin(S)\,(\cT^{-1})^\dagger\cT^{-1}\le\one\) because \([(\cT^{-1})^\dagger\cT^{-1}]_{\rm L}=S^{-1}\), and maps \(\ket{\Psi_\vphi}\) to \(\sqrt{\lmin(S)/\cN}\,\ket{\Phi_\vphi}\); followed by the incoherent unitary \(\ket{l_j}\mapsto e^{-i\varphi_j}\ket j\), it succeeds with the optimal probability \(\lmin(S)/\cN=d\lmin(S)\mu(\Psi_\vphi)\).

\textit{Deterministic sources for equal overlaps.} By the first statement, \(\Pst=1\) iff \(G_\psi\ge\one/d\), that is, since \(X\) is invertible, iff \(S\ge\diag(|x_j|^{-2})/d\). The projector onto the eigenspace of \(\lmin(S)\) is \(J/d\) for \(s<0\) and \(\one-J/d\) for \(s>0\); in both cases all its diagonal entries are equal. The trace of the inequality against this projector therefore gives \(\lmin(S)\ge d^{-2}\sum_j|x_j|^{-2}\). On the other hand, \(\sum_j|x_j|^{-2}\ge d^2/\|x\|^2\ge d^2\lmin(S)\), by the inequality between the arithmetic and harmonic means and by \(1=x^\dagger Sx\ge\lmin(S)\|x\|^2\). Equality must therefore hold throughout, which requires equal moduli and \(x\) in the eigenspace of \(\lmin(S)\), that is, a \(\mu\)-maximal state [Proposition~\ref{prop:mu}(c)]. Conversely, every \(\mu\)-maximal state is balanced with \(\cN=\lmin(S)\), so that \(\Pst=\lmin(S)/\cN=1\).

(ii) The source has \(G_\psi=\one/d\), which is permutation invariant with \(\tau=1\), so Theorem~\ref{thm:inphase}(b) gives \(\Pst=\min\{1,1/\tau(\phi)\}\), and \(\tau=(d-\cN)/[(d-1)\cN]\) for balanced targets. The operator \(\cT/\sqrt{\lmax(S)}\) is superposition-free from \(\{\ket{l_j}\}\) to \(\{\ket{c_j}\}\), satisfies \(\cT^\dagger\cT/\lmax(S)\le\one\), and maps \(\ket{\Phi_\vphi}\) to a multiple of \(\ket{\Psi_\vphi}\) with squared norm \(\cN/\lmax(S)\le1\). It is optimal iff \(\cN/\lmax(S)=\min\{1,(d-1)\cN/(d-\cN)\}\), that is, iff \(\cN=\lmax(S)\), where it succeeds with certainty, or \(\cN=d-(d-1)\lmax(S)<1\). For equal overlaps with \(s>0\), \(\lmax(S)=\lambda_0\): the first case is \(\ket{\Psi_0}\), and the second requires \(\cN=1-(d-1)^2s\), which is compatible with \(\cN\ge\lambda_1=1-s\) only for \(d=2\), where it gives \(\cN=\lambda_1\), the state \(\ket{\Psi_-}\). For \(s<0\), \(\lmax(S)=\lambda_1\): the first case is class (A), and the second gives \(\cN=\lambda_0\), the state \(\ket{\Psi_0}\).

(iii) Let \(\tilde s=s/\lambda_1(s)\) and \(\tilde s'=s'/\lambda_1(s')\); \(\tilde s\) increases with \(s\), and \(1/\tau_0=1+d\tilde s\) and \(\tau_A=1+d\tilde s/(d-1)\). Equation~\eqref{eq:kappa} is Theorem~\ref{thm:inphase}(b). For Eq.~\eqref{eq:eta}, write \(\tau_A'=\tau_A(s')\). By Eq.~\eqref{eq:GclassA}, \(dG_\psi=\one+\tilde s\,vv^\dagger\) and \(d\,\Pi_\pi^TG_\phi\Pi_\pi=\one+\tilde s'v_\pi v_\pi^\dagger\) with \(v_\pi=\Pi_\pi^Tv\); both \(v\) and \(v_\pi\) are orthogonal to \(\ones\) and have squared norm \(d\). The upper bound \(\tau_A(s)/\tau_A'\) is Theorem~\ref{thm:inphase}(a). If \(s'\le s\), the identity branch alone satisfies the constraint with probability one. If \(0\le s<s'\), take the identity branch with probability \(\tilde s/\tilde s'\) and a uniform twirl over a doubly transitive group with total probability \((1-\tilde s/\tilde s')/\tau_A'\). By Eq.~\eqref{eq:twirlG}, the twirl contributes \((1-\tilde s/\tilde s')[J/(d\tau_A')+\one-J/d]\) to \(d\) times the left-hand side of the constraint, which then holds with equality on \(v\) and on \(\{\ones,v\}^\perp\) and strictly on \(\ones\). The two probabilities add up to \(\tau_A(s)/\tau_A'\). For \(d=2\), Eq.~\eqref{eq:eta} holds for all \(s\) and \(s'\) by Theorem~\ref{thm:inphase}(b), because \(G_{\Psi_-}\) is permutation invariant. Finally, let \(s<s'\le0\), so that \(\tilde s<\tilde s'\le0\). The identity branch with probability \((1+d\tilde s)/(1+d\tilde s')=\tau_0(s')/\tau_0(s)<1\) satisfies the constraint, with equality on \(v\). Conversely, the expectation value of the constraint in \(v\) gives \(\sum_\pi w_\pi(d+\tilde s'|v_\pi^\dagger v|^2)\le d\,(1+d\tilde s)\), and \(|v_\pi^\dagger v|^2\le d^2\) with \(\tilde s'\le0\) gives \(d+\tilde s'|v_\pi^\dagger v|^2\ge d\,(1+d\tilde s')\), so that \(\sum_\pi w_\pi\le\tau_0(s')/\tau_0(s)\).
\end{proof}

\section{Numerical checks}\label{app:numerics}

The semidefinite program~\eqref{eq:primal} was solved numerically with CVXPY~\cite{CVXPY2016} and the Clarabel interior-point solver~\cite{Clarabel2024} after whitening by \(G_\psi^{-1/2}\). Every closed form, every entry of Tables~\ref{tab:cats}--\ref{tab:roles}, and every value in Figs.~\ref{fig:graphs} and~\ref{fig:qubit} agrees with it to within \(10^{-7}\), on random instances with \(d=2,\dots,5\) and both signs of \(s\), on random pairs of source and target reference sets with \(d\le4\) (Theorem~\ref{thm:inphase}), and on the circulant Gram matrices of the coherent states \(\ket{\alpha\omega^j}\) (Lemma~\ref{lem:fourier}).

Further statements were checked individually: Proposition~\ref{prop:mu}(a) and the monotonicity of \(\tau\) on 400 random stochastic FO each; Eq.~\eqref{eq:secondorder} for \(s=0.01\)--\(0.04\); Proposition~\ref{prop:copies} against the program with all 24 permutations for \(N=2\) and all \(40320\) for \(N=3\), where the solver reproduces \(r^4\) to within \(3\times10^{-7}\); Proposition~\ref{prop:qubit} on 600 random complex sources and targets at five values of \(s\) of both signs; Corollary~\ref{cor:copies} for two copies; Corollary~\ref{cor:phase} on a \(25\times25\) grid of phases at four values of \(s\); Theorems~\ref{thm:nogo}(i) and~\ref{thm:price} with Fourier and non-Fourier class-(A) sources for \(d=3,4,5\) and random targets; the worst case \(1/\tau_A\) of Theorem~\ref{thm:price} by global searches over all full-support targets, for \(d=3\) at three values of \(s\) and for \(d=4\) with a non-Fourier class-(A) source; Eq.~\eqref{eq:qutritall} on 400 random, real, and nearly chiral targets at five values of \(s\); the statement on two-component targets through the certificate~\eqref{eq:Yprice} for \(d=3,\dots,6\), including a non-Fourier pattern with \(\sum_je^{2i\varphi_j}=0\) for \(d=6\), and through the exact program near these targets; and Theorem~\ref{thm:bridge}(i) on 4000 random sources per dimension and sign of \(s\) and Theorem~\ref{thm:bridge}(iii) for all sign combinations of \(s\) and \(s'\), with Fourier and non-Fourier class-(A) states for \(d=2,\dots,5\).

Three results are numerical. The means for random sources in Sec.~\ref{sec:sdp} use 1500 Haar-random sources for \(d=3\) and 500 for \(d=4\), drawn with a fixed seed; their standard errors are at most \(0.005\). The remark after Theorem~\ref{thm:price} comes from a global search over all full-support targets. For \(N=4\) copies, an exhaustive search over the \(20160\) invertible binary matrices \(A\) finds \(\max_{k\neq0}(|k|+|Ak|)\ge6\) for each of them, so no single linear branch attains \(r^5\); a linear program over these branches shows that their mixtures attain \(r^5\) for \(r\ge0.225\), in exact rational arithmetic for \(r=\tfrac12\) and \(r=\tfrac13\), but not for \(r\le0.224\).

\bibliography{references}

\end{document}